\documentclass[aip,jcp,preprint,showkeys]{revtex4-1}
\usepackage{graphicx,dcolumn,bm,color,microtype,hyperref,multirow,amscd,amsmath,amssymb,amsfonts}

\newcommand{\alert}[1]{\textcolor{black}{#1}}
\newcommand{\eps}{\varepsilon}
\newcommand{\lam}{\lambda}
\newcommand{\mc}{\multicolumn}
\newcommand{\mr}{\multirow}
\newcommand{\et}{e_\text{t}}
\newcommand{\ex}{e_\text{x}}
\newcommand{\ec}{e_\text{c}^\text{FF}}
\newcommand{\eFF}{e^\text{FF}}
\newcommand{\eWC}{e^\text{WC}}
\newcommand{\eHF}{e_\text{HF}^\text{FF}}

\newcommand{\ku}{k_{\uparrow}}
\newcommand{\kd}{k_{\downarrow}}

\newcommand{\kF}{k_{\text{F}}}
\newcommand{\upsa}{\Upsilon_0^{\text{a}}}
\newcommand{\upsb}{\Upsilon_0^{\text{b}}}
\newcommand{\Lama}{\Lambda_1^{\text{a}}}
\newcommand{\Lamb}{\Lambda_1^{\text{b}}}
\newcommand{\br}{\bm{r}}

\newcommand{\rs}{r_s}
\newcommand{\cdash}{\multicolumn{1}{c}{---}}
\DeclareMathOperator{\Li}{Li}	
\DeclareMathOperator{\erfc}{erfc}

\begin{document}	

\title{The uniform electron gas}	

\author{Pierre-Fran{\c c}ois Loos}
\email{pf.loos@anu.edu.au}
\thanks{Corresponding author}
\affiliation{Research School of Chemistry, Australian National University, Canberra ACT 2601, Australia}
\author{Peter M. W. Gill}
\email{peter.gill@anu.edu.au}
\affiliation{Research School of Chemistry, Australian National University, Canberra ACT 2601, Australia}

\begin{abstract}
The uniform electron gas or UEG (also known as jellium) is one of the most fundamental models in condensed-matter physics and the cornerstone of the most popular approximation --- the local-density approximation --- within density-functional theory.  In this article, we provide a detailed review on the energetics of the UEG at high, intermediate and low densities, and in one, two and three dimensions.  We also report the best quantum Monte Carlo and symmetry-broken Hartree-Fock calculations available in the literature for the UEG and discuss the phase diagrams of jellium.
\end{abstract}


\maketitle

%----------------------------------------------------------------
\section{Introduction}
%----------------------------------------------------------------
The final decades of the twentieth century witnessed a major revolution in solid-state and molecular physics, as the introduction of sophisticated exchange-correlation models \cite{ParrBook} propelled density-functional theory (DFT) from qualitative to quantitative usefulness.  The apotheosis of this development was probably the award of the 1998 Nobel Prize for Chemistry to Walter Kohn \cite{Kohn99} and John Pople \cite{Pople99} but its origins can be traced to the prescient efforts by Thomas, Fermi and Dirac, more than 70 years earlier, to understand the behavior of ensembles of electrons without explicitly constructing their full wave functions.

In principle, the cornerstone of modern DFT is the Hohenberg-Kohn theorem \cite{Hohenberg64} but, in practice, it rests largely on the presumed similarity between the electronic behavior in a real system and that in the hypothetical \alert{three-dimensional} uniform electron gas (UEG). \cite{VignaleBook}  
\alert{This model system was applied by Sommerfeld in the early days of quantum mechanics to study metals \cite{Sommerfeld28} and}
in 1965, Kohn and Sham \cite{Kohn65} showed that the knowledge of a analytical parametrization of the UEG correlation energy allows one to perform approximate calculations for atoms, molecules and solids.  This spurred the development of a wide variety of spin-density correlation functionals (VWN, \cite{Vosko80} PZ, \cite{Perdew81} PW92, \cite{Perdew92} \textit{etc.}), each of which requires information on the high- and low-density regimes of the spin-polarized UEG, and are parametrized using numerical results from Quantum Monte Carlo (QMC) calculations, \cite{Ceperley78, Ceperley80} together with analytic perturbative results.

For this reason, a detailed and accurate understanding of the properties of the UEG \alert{ground state} is essential to underpin the continued evolution of DFT.  
\alert{Moreover, meaningful comparisons between theoretical calculations on the UEG and realistic systems (such as sodium) have also been performed recently (see, for example, Ref.~\onlinecite{Huotari10}).}
\alert{The two-dimensional version of the UEG has also been the object of extensive research \cite{Ando82, Abrahams01} because of its intimate connection to two-dimensional or quasi-two-dimensional materials, such as quantum dots. \cite{Alhassid00, Reimann02}
The one-dimensional UEG has recently attracted much attention due to its experimental realization in carbon nanotubes, \cite{SaitoBook, Egger98, Bockrath99, Ishii03, Shiraishi03} organic conductors, \cite{Schwartz98, Vescoli00, Lorenz02, Dressel05, Ito05} transition metal oxides, \cite{Hu02} edge states in quantum Hall liquids, \cite{Milliken96, Mandal01, Chang03} semiconductor heterostructures, \cite{Goni91, Auslaender00, ZaitsevZotov00, Liu05, Steinberg06} confined atomic gases, \cite{Monien98, Recati03, Moritz05} and atomic or semiconducting nanowires. \cite{Schafer08, Huang01}}
In the present work, we have attempted to collect and collate the key results on the energetics of the UEG, information that is widely scattered throughout the physics and chemistry literature.  Section \ref{sec:UEG} defines and describes the UEG model in detail.  Section \ref{sec:HDL} reports the known results for the high-density regime, wherein the UEG is a Fermi fluid (FF) of delocalized electrons.  Section \ref{sec:LDL} reports analogous results for the low-density regime, in which the UEG becomes a Wigner crystal (WC) of relatively localized electrons.  The intermediate-density results from QMC and symmetry-broken Hartree-Fock (SBHF) calculations are gathered in Sec.~\ref{sec:IDR}.  Atomic units are used throughout.

%----------------------------------------------------------------
\section{
\label{sec:UEG}
UEG paradigm}
%----------------------------------------------------------------
The $D$-dimensional uniform electron gas, or $D$-jellium, consists of interacting electrons in an infinite volume in the presence of a uniformly distributed background of positive charge.  
Traditionally, the system is constructed by allowing the number $n=n_{\uparrow}+n_{\downarrow}$ of electrons (where $n_{\uparrow}$ and $n_{\downarrow}$ are the numbers of spin-up and spin-down electrons, respectively) in a $D$-dimensional cube of volume $V$ to approach infinity with the density $\rho = n/V$ held constant. \cite{ParrBook}
The spin polarization is defined as
\begin{equation}
	\zeta = \frac{\rho_{\uparrow} - \rho_{\downarrow}}{\rho} = \frac{n_{\uparrow} - n_{\downarrow}}{n},
\end{equation}
where ${\rho_\uparrow}$ and ${\rho_\downarrow}$ is the density of the spin-up and spin-down electrons, respectively, and the $\zeta = 0$ and $\zeta = 1$ cases are called paramagnetic and ferromagnetic UEGs.

The total ground-state energy of the UEG (including the positive background) is
\begin{equation}
	E[\rho] = T_{\text{s}}[\rho] + \int \rho(\br) v(\br) d\br + J[\rho] + E_\text{xc}[\rho] + E_\text{b},
\end{equation}
where $T_{\text{s}}$ is the non-interacting kinetic energy, 
\begin{equation}
	v(\br) = - \int \frac{\rho_\text{b}(\br^\prime)}{\left| \br - \br^\prime \right|} d\br^\prime
\end{equation}
is the external potential due to the positive background density $\rho_\text{b}$, 
\begin{equation}
	J[\rho] = \frac{1}{2} \iint \frac{\rho(\br)\rho(\br^\prime)}{\left| \br - \br^\prime \right|} d\br d\br^\prime
\end{equation}
is the Hartree energy, $E_{\text{xc}}$ is the exchange-correlation energy and
\begin{equation}
	E_\text{b} = \frac{1}{2} \iint \frac{\rho_\text{b}(\br)\rho_\text{b}(\br^\prime)}{\left| \br - \br^\prime \right|} d\br d\br^\prime
\end{equation}
is the electrostatic self-energy of the positive background.
The neutrality of the system [$\rho(\br) = \rho_\text{b}(\br)$] implies that 
\begin{equation}
	\int \rho(\br) v(\br) d\br + J[\rho] + E_\text{b} = 0,
\end{equation}
which yields
\begin{equation}
\label{eq:E-jellium}
\begin{split}
	E[\rho]	& = T_{\text{s}}[\rho] + E_\text{xc}[\rho] 
	\\
			& = T_{\text{s}}[\rho] + E_\text{x}[\rho] + E_\text{c}[\rho] 
	\\
			& = \alert{\int \rho\,\et[\rho] d\br + \int \rho\,\ex[\rho] d\br + \int \rho\,e_\text{c}[\rho] d\br.}
\end{split}
\end{equation}
In the following, we will focus on the three reduced (i.e.~per electron) energies $\et$, $\ex$ and $e_\text{c}$ and we will discuss these as functions of the Wigner-Seitz radius $\rs$ defined via
\begin{equation}
	\frac{1}{\rho} = 
	\frac{\pi^{D/2}}{\Gamma\left( \frac{D}{2} + 1\right)} \rs^D
	=
	\begin{cases}
		\frac{4\pi}{3} \rs^3,	&	D = 3,	\\
		\pi\,\rs^2,			&	D = 2,	\\
		2\,\rs,				&	D = 1,	\\
	\end{cases} 
\end{equation}
or
\begin{equation}
	\rs = 
	\begin{cases}
		\left(\frac{3}{4\pi\rho}\right)^{1/3},	&	D = 3,	\\
		\left(\frac{1}{\pi\rho}\right)^{1/2},		&	D = 2,	\\
		\frac{1}{2\rho},						&	D = 1,	\\
	\end{cases} 
\end{equation}
where $\Gamma$ is the Gamma function. \cite{NISTbook}  It is also convenient to introduce the Fermi wave vector  
\begin{equation}
	\kF = \frac{\alpha}{\rs},
\end{equation}
where
\begin{equation}
\label{eq:alpha}
	\alpha = 2^{\frac{D-1}{D}} \Gamma\left( \frac{D}{2} + 1\right)^{2/D}
	=
	\begin{cases}
		\left(\frac{9\pi}{4}\right)^{1/3},		&	D = 3,	\\
		\sqrt{2},						&	D = 2,	\\
		\frac{\pi}{4},					&	D = 1.	\\
	\end{cases} 
\end{equation}

%----------------------------------------------------------------
\section{
\label{sec:HDL}
The high-density regime}
%----------------------------------------------------------------
In the high-density regime ($\rs \ll 1$), also called the weakly-correlated regime, the kinetic energy of the electrons dominates the potential energy, resulting in a completely delocalized system. \cite{VignaleBook}  In this regime, \alert{ the one-electron orbitals are plane waves and} the UEG is described as a Fermi fluid (FF). Perturbation theory yields the energy expansion
\begin{equation}
\label{eq:def-eFF}
	\eFF(\rs,\zeta) = \et(\rs,\zeta) + \ex(\rs,\zeta) + \ec(\rs,\zeta),
\end{equation}
where the non-interacting kinetic energy $\et(\rs,\zeta)$ and exchange energy $\ex(\rs,\zeta)$ are the zeroth- and first-order perturbation energies, respectively, and the correlation energy $\ec(\rs,\zeta)$ encompasses all higher orders.

%----------------------------------------------------------------
\subsection{Non-interacting kinetic energy}
%----------------------------------------------------------------
The non-interacting kinetic energy of $D$-jellium is the first term of the high-density energy expansion \eqref{eq:def-eFF}.  The 3D case has been known since the work of Thomas and Fermi \cite{Thomas27, Fermi26} and, for $D$-jellium, it reads \cite{Glasser83, Iwamoto84}
\begin{equation}
\label{eq:et}
	\et(\rs,\zeta) = \frac{\eps_\text{t}(\zeta)}{\rs^2},
\end{equation}
where
\begin{subequations} 
\begin{gather}
	\eps_\text{t}(\zeta) = \eps_\text{t} \Upsilon_\text{t}(\zeta),
	\\
	\label{eq:epst}
	\eps_\text{t} \equiv \eps_\text{t}(\zeta=0) = \frac{D}{2(D+2)} \alpha^2,
\end{gather}
\end{subequations} 
and the spin-scaling function is
\begin{equation}
	\label{eq:SPF-t}
	\Upsilon_\text{t}(\zeta) = \frac{\left(1+\zeta\right)^{\frac{D+2}{D}}+\left(1-\zeta\right)^{\frac{D+2}{D}}}{2}.
\end{equation}
The values of $\eps_\text{t}(\zeta)$ in the paramagnetic and ferromagnetic limits are given in Table \ref{tab:HDL} for $D=1$, $2$ and $3$.

%----------------------------------------------------------------
\subsection{Exchange energy}
%----------------------------------------------------------------
The exchange energy, which is the second term in \eqref{eq:def-eFF}, can be written \cite{Dirac30, Friesecke97}
\begin{equation}
\label{eq:ex}
	\ex(\rs,\zeta) = \frac{\eps_\text{x}(\zeta)}{\rs},
\end{equation}
where
\begin{subequations} 
\begin{gather}
	\eps_\text{x}(\zeta) = \eps_\text{x} \Upsilon_\text{x}(\zeta),
	\\
	\eps_\text{x} \equiv \eps_\text{x}(\zeta=0) = -\frac{2D}{\pi(D^2-1)} \alpha,
	\label{eq:epsx}
	\\
	\label{eq:SPF-x}
	\Upsilon_\text{x}(\zeta) = \frac{\left(1+\zeta\right)^{\frac{D+1}{D}}+\left(1-\zeta\right)^{\frac{D+1}{D}}}{2}.
\end{gather}
\end{subequations} 
The values of $\eps_\text{x}(\zeta)$ in the paramagnetic and ferromagnetic limits are given in Table \ref{tab:HDL} for $D=1$, $2$ and $3$.
\alert{Note that, due to the particularly strong divergence of the Coulomb operator, $\eps_\text{x}(\zeta)$ diverges in 1D.}

%%% TABLE 1 %%%
\begin{turnpage}
\begin{table*}
\caption{
\label{tab:HDL}
Energy coefficients for the paramagnetic ($\zeta=0$) and ferromagnetic ($\zeta=1$) states and spin-scaling functions of $D$-jellium at high density.
Note that $\gamma$ is the Euler-Mascheroni constant, $z(n)$ is the Riemann zeta function and $\beta$ is the Dirichlet beta function. \cite{NISTbook}
In 1-jellium, the paramagnetic and ferromagnetic states are degenerate.}
\begin{ruledtabular}
\begin{tabular}{ccccccccccc}
	\mr{2}{*}{Term}		&	\mr{2}{*}{Coefficient}			&	\mc{3}{c}{Paramagnetic state}				&	\mc{3}{c}{Ferromagnetic state}					&	\mc{3}{c}{Spin-scaling function}								\\
					&							&	\mc{3}{c}{$\eps(0)$, $\lam(0)$}				&	\mc{3}{c}{$\eps(1)$, $\lam(1)$}					&	\mc{3}{c}{$\Upsilon(\zeta)$, $\Lambda(\zeta)$}					\\
													\cline{3-5}								\cline{6-8}										\cline{9-11}
					&							&	$D = 1$	&	$D = 2$	&	$D = 3$		&	$D = 1$	&	$D = 2$	&	$D = 3$			&	$D = 1$	&	$D = 2$	&	$D = 3$						\\
\hline		
	$\rs^{-2}$			&	$\eps_\text{t}(\zeta)$			&	$\pi^2/96$								&	$1/2$									&	$\frac{3}{10} \left(\frac{9\pi}{4}\right)^{2/3}$			
												&	$\pi^2/24$								&	$1$										&	$2^{2/3} \frac{3}{10} \left(\frac{9\pi}{4}\right)^{2/3}$			
												&	$1$									&	Eq.~\eqref{eq:SPF-t}							&	Eq.~\eqref{eq:SPF-t}										\\[8pt]
	$\rs^{-1}$			&	$\eps_\text{x}(\zeta)$		&	$-\infty$								&	$-\frac{4\sqrt{2}}{3\pi}$						&	$-\frac{3}{4\pi} \left(\frac{9\pi}{4}\right)^{1/3}$		
												&	$-\infty$								&	$-\frac{8}{3\pi}$								&	$- 2^{1/3} \frac{3}{4\pi} \left(\frac{9\pi}{4}\right)^{1/3}$			
												&	$1$									&	Eq.~\eqref{eq:SPF-x}						&	Eq.~\eqref{eq:SPF-x}									\\[8pt]
	$\ln \rs$			&	$\lam_{0}(\zeta)$			&	$0$									&	$0$										&	$\frac{1-\ln2}{\pi^2}$		
												&	$0$									&	$0$										&	$\frac{1-\ln2}{2\pi^2}$			
												&	\cdash								&	\cdash									&	Eq.~\eqref{eq:L0-3D}									\\[8pt]
	$\rs^0$			&	$\eps_0^\text{a}(\zeta)$		&	$-\pi^2/360$							&	$\ln 2 - 1$									&	$-0.071\,100$				
												&	$-\pi^2/360$							&	$\frac{\ln 2 - 1}{2}$							&	$-0.049\,917$			
												&	$1$									&	Eq.~\eqref{eq:E0a-2D}						&	Ref.~\onlinecite{Hoffman92}									\\[8pt]
					&	$\eps_0^\text{b}(\zeta)$		&	$0$									&	$\beta(2) - \frac{8}{\pi^2}\beta(4)$				&	$\frac{\ln 2}{6} - \frac{3}{4\pi^2} z(3)$						
												&	$0$									&	$\beta(2) - \frac{8}{\pi^2}\beta(4)$				&	$\frac{\ln 2}{6} - \frac{3}{4\pi^2} z(3)$		
												&	\cdash								&	$1$										&	$1$													\\[8pt]
	$\rs\ln \rs$			&	$\lam_1^\text{a}(\zeta)$		&	$0$									&	$-\sqrt{2}\left(\frac{10}{3\pi}-1\right)$				&	$\left(\frac{9\pi}{4}\right)^{1/3} \frac{\pi^2-6}{24\pi^3}$
												&	$0$									&	$-\frac{1}{4}\left(\frac{10}{3\pi}-1\right)$			&	$\frac{1}{2^{7/3}} \left(\frac{9\pi}{4}\right)^{1/3}  \frac{\pi^2+6}{24\pi^3}$
												&	\cdash								&	Eq.~\eqref{eq:L1-2D}						&	Eq.~\eqref{eq:L1a-3D}									\\[8pt]
					&	$\lam_1^\text{b}(\zeta)$		&	$0$									&	$0$										&	$\left(\frac{9\pi}{4}\right)^{1/3}  \frac{\pi^2-12\ln 2}{4\pi^3}$
												&	$0$									&	$0$										&	 $\frac{1}{2^{4/3}} \left(\frac{9\pi}{4}\right)^{1/3}  \frac{\pi^2-12\ln 2}{4\pi^3}$
												&	\cdash								&	---										&	Eq.~\eqref{eq:L1b-3D}									\\[8pt]
	$\rs$				&	$\eps_1(\zeta)$				&	$+0.008\,446$							&	unknown									&	$-0.010$		
												&	$+0.008\,446$							&	unknown									&	unknown
												&	$1$									&	unknown									&	unknown												\\[8pt]
\end{tabular}
\end{ruledtabular}
\end{table*}
\end{turnpage}

%----------------------------------------------------------------
\subsection{Hartree-Fock energy}
%----------------------------------------------------------------
In the high-density limit, one might expect the Hartree-Fock (HF) energy of the UEG to be the sum of the kinetic energy \eqref{eq:et} and the exchange energy \eqref{eq:ex}, i.e.
\begin{equation}
\label{eq:eHF}
	\eHF(\rs,\zeta) = \et(\rs,\zeta) + \ex(\rs,\zeta).
\end{equation}
However, although this energy corresponds to a solution of the HF equation, a stability analysis\cite{VignaleBook} reveals that \eqref{eq:eHF} is never the lowest possible HF energy and Overhauser showed \cite{Overhauser59, Overhauser62} that it is always possible to find a symmetry-broken solution of lower energy.  We will discuss this further in Sec.~\ref{sec:SBHF}.

In 1D systems, the Coulomb operator is so strongly divergent that a new term appears in the HF energy expression.  Thus, for 1-jellium, Fogler found\cite{Fogler05a} that
\begin{equation}
\label{eq:eHF-1D}
	\eHF(\rs) = \frac{\pi^2}{24 \rs^2} - \frac{1}{2} \frac{\ln \rs}{\rs} + \frac{2 \ln(\pi/2) - 3 + 2 \gamma}{4\rs}, 
\end{equation}
where $\gamma$ is the Euler-Mascheroni constant. \cite{NISTbook}  Furthermore, because the paramagnetic and ferromagnetic states are degenerate in strict 1D systems, we can confine our attention to the latter. \cite{Astrakharchik11, Lee11a, QR12, Ringium13, gLDA14, 1DChem15}

%----------------------------------------------------------------
\subsection{Correlation energy}
%----------------------------------------------------------------
The high-density correlation energy expansions 
\begin{equation}
	\ec(\rs,\zeta) = e(\rs,\zeta) - \eHF(\rs,\zeta)
\end{equation}
of the two- and three-dimensional UEGs have been well studied. \cite{Zia73, Isihara77, Rajagopal77, Glasser77, Isihara80, Glasser84, Seidl04, Chesi07, 2DEG11, Macke50, Bohm53, Pines53, GellMann57, DuBois59, Carr64, Misawa65, Onsager66, Wang91, Hoffman92, Endo99, Ziesche05, Sun10, 3DEG11, Handler12, Glomium11}  Much less is known about 1-jellium. \cite{Ringium13, 1DEG13}  \alert{Using Rayleigh-Schr\"odinger perturbation theory,} the correlation energy seems to possess the expansion 
\begin{align}
\label{eq:EcHDL}
	\ec(\rs,\zeta)	& = \sum_{j=0}^\infty \left[\lam_j(\zeta) \ln \rs + \eps_j(\zeta) \right] \rs^j	\notag	\\
					& = \lam_0(\zeta) \ln \rs + \eps_0(\zeta)									\notag	\\
					&+ \lam_1(\zeta) \rs \ln \rs + \eps_1(\zeta) \rs + \ldots
\end{align}
and the values of these coefficients (when known) are given in Table \ref{tab:HDL}.  The methods for their determination are outlined in the next three subsections.

%*************************
\subsubsection*{3-jellium}
%*************************
The coefficient $\lam_0(\zeta)$ can been obtained by the Gell-Mann--Brueckner resummation technique, \cite{GellMann57} which sums the most divergent terms of the series \eqref{eq:EcHDL} to obtain
\begin{equation}
	\lam_0(\zeta) = \frac{3}{32\pi^3} \int_{-\infty}^{\infty} \left[R_0(u,\zeta)\right]^2 du,
\end{equation}
where
\begin{subequations} 
\begin{gather}
	\label{eq:R0u-zeta}
	R_0(u,\zeta) = \kd R_0\left(\frac{u}{\kd}\right) + \ku R_0\left(\frac{u}{\ku}\right),
	\\
	\label{eq:R0u}
	R_0(u) = 1 - u \arctan(1/u),
\end{gather}
\end{subequations} 
and 
\begin{equation}
\label{eq:kupdown}
	k_{\uparrow,\downarrow} = \left(1\pm\zeta\right)^{1/D}
\end{equation}
is the Fermi wave vector of the spin-up or -down electrons.  

The paramagnetic \cite{Macke50} and ferromagnetic \cite{Misawa65} limits are given in Table \ref{tab:HDL}, and the spin-scaling function
\begin{equation} 
\label{eq:L0-3D}
	\Lambda_0(\zeta) 	= \frac{1}{2} + \frac{1}{4(1-\ln 2)} \bigg[\kd \ku (\kd+\ku)	
%	\\
						- \kd^3 \ln \left(1+\frac{\ku}{\kd}\right) - \ku^3 \ln \left(1+\frac{\kd}{\ku}\right) \bigg]
\end{equation}
was obtained by Wang and Perdew. \cite{Wang91}

The coefficient $\eps_0(\zeta)$ is often written as the sum
\begin{equation} 
\label{eq:eps0}
	\eps_0(\zeta) = \eps_{0}^{\text{a}}(\zeta) + \eps_{0}^{\text{b}}(\zeta)
\end{equation}
of a RPA (random-phase approximation) or ``ring-diagram'' term $\eps_{0}^{\text{a}}(\zeta)$ and a first-order exchange term $\eps_0^{\text{b}}(\zeta)$.  
The RPA term $\eps_{0}^{\text{a}}(\zeta)$ is not known in closed form but it can be computed numerically with high precision. \cite{Hoffman92}
Its paramagnetic and ferromagnetic limits are given in Table \ref{tab:HDL} and the spin-scaling function 
\begin{equation}
	\upsa(\zeta) = \eps_0^{\text{a}}(\zeta)/\eps_0^{\text{a}}(0)
\end{equation}
can be found using Eq.~(20) in Ref.~\onlinecite{Hoffman92}.
The first-order exchange term \cite{Onsager66} is given in Table \ref{tab:HDL} and, because it is independent of the spin-polarization, the spin-scaling function 
\begin{equation}
\upsb(\zeta) = \eps_0^{\text{b}}(\zeta)/\eps_0^{\text{b}}(0) = 1
\end{equation}
is trivial.

The coefficient $\lam_1(\zeta)$ can be written similarly \cite{Carr64} as 
\begin{equation}
	\lam_1(\zeta) = \lam_{1}^{\text{a}}(\zeta) + \lam_{1}^{\text{b}}(\zeta),
\end{equation}
where
\begin{subequations} 
\begin{gather}
	\lam_{1}^{\text{a}}(\zeta) = -\frac{3\alpha}{8\pi^5} \int_{-\infty}^{\infty} \mathcal{R}_1^{\text{a}}(u,\zeta) \,du,	
	\label{eq:lam1a-zeta}	
	\\
	\lam_{1}^{\text{b}}(\zeta) = \frac{3\alpha}{16\pi^4} \int_{-\infty}^{\infty} \mathcal{R}_1^{\text{b}}(u,\zeta) \,du	
	\label{eq:lam1b-zeta}
\end{gather}
\end{subequations} 
are the RPA and second-order exchange contributions and $\alpha$ is given in \eqref{eq:alpha}.
The integrands are \cite{Perdew92, Sun10}
\begin{subequations}
\begin{gather}
	\mathcal{R}_1^{\text{a}}(u,\zeta) = R_0(u,\zeta)^2 R_1(u,\zeta),	\\
	\mathcal{R}_1^{\text{b}}(u,\zeta) = R_0(u,\zeta) R_2(i u,\zeta),	\\
	\label{eq:R1u-zeta}
	R_1(u,\zeta) = \kd^{-1} R_1\left(\frac{u}{\kd}\right) + \ku^{-1} R_1\left(\frac{u}{\ku}\right),	
	\\
	R_2(i u,\zeta) = R_2\left(i\frac{u}{\kd}\right) + R_2\left(i\frac{u}{\ku}\right),
	\\
	R_1(u) = -\frac{\pi}{3(1+u^2)^2},					\\
	R_2(i u) = 4\frac{(1+3u^2) - u(2+3u^2) \arctan u}{1+u^2}.
\end{gather}
\end{subequations}
Carr and Maradudin gave an estimate \cite{Carr64} of $\lam_1(0)$ and this was later refined by Perdew and coworkers. \cite{Perdew92, Sun10}

However, we have found \cite{3DEG11} that the integrals in Eqs.~\eqref{eq:lam1a-zeta} and \eqref{eq:lam1b-zeta} can be evaluated exactly by computer software, \cite{Mathematica7} giving the paramagnetic and ferromagnetic values in Table \ref{tab:HDL} and the spin-scaling functions
\begin{subequations} 
\begin{align} 
\begin{split} 
	\label{eq:L1a-3D}
	\Lambda_1^{\text{a}}(\zeta) 
						& = \frac{3}{\pi^2-6} \left\{ \left(\frac{\pi^2}{6}+\frac{1}{4}\right)(\kd^2+\ku^2) \right.
						\\
						& - \frac{3}{2}\kd\ku -\frac{\kd^2+\ku^2}{\kd^2-\ku^2}\kd\ku\ln\left(\frac{\kd}{\ku}\right)
						\\
						& \left. -\frac{\kd^2-\ku^2}{2}\left[\Li_2\left(\frac{\kd-\ku}{\kd+\ku}\right) - \Li_2\left(\frac{\ku-\kd}{\kd+\ku}\right) \right]\right\},
\end{split}
\\
\begin{split} 
	\label{eq:L1b-3D}
	\Lambda_1^{\text{b}}(\zeta) 
						& = \frac{3}{\pi^2-12\ln2} \left\{ \frac{\pi^2}{6}(\kd^2+\ku^2) \right. 
						\\
						& + (1-\ln2)(\kd-\ku)^2 - \frac{\kd^2}{2}\Li_2\left(\frac{\kd-\ku}{\kd+\ku}\right) 	
						\\
						& - \frac{\ku^2}{2}\Li_2\left(\frac{\ku-\kd}{\kd+\ku}\right) + \frac{1}{\kd\ku} \left[\kd^4\ln\left(\frac{\kd}{\kd+\ku}\right)  \right.
						\\
						&  \left.  \left. + \kd^2\ku^2\ln\left(\frac{\kd\ku}{(\kd+\ku)^2}\right) + \ku^4\ln\left(\frac{\ku}{\kd+\ku}\right) \right] \right\},
\end{split}
\end{align}
\end{subequations}
where $\Li_2$ is the dilogarithm function. \cite{NISTbook}

\begin{figure}
\includegraphics[width=0.45\textwidth]{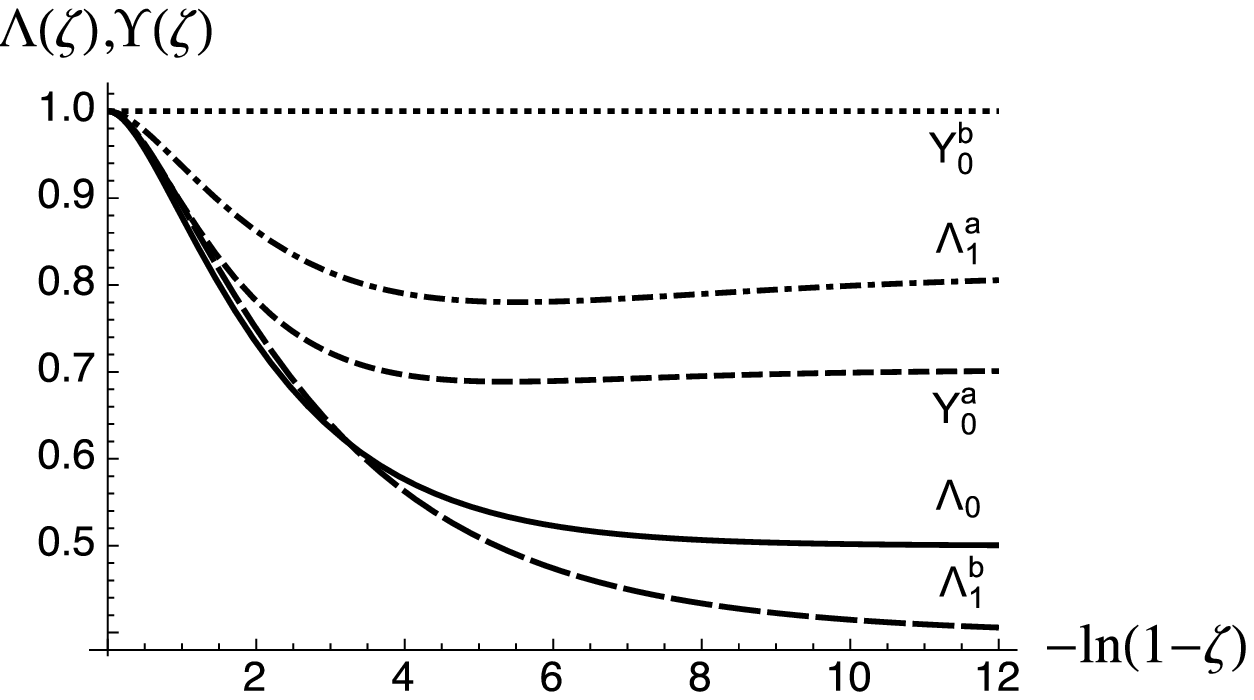}
\caption{
\label{fig:3jellium-HDL-SSF}
Spin-scaling functions of 3-jellium as functions of $\zeta$.}
\end{figure}

The spin-scalings $\Lambda_0(\zeta)$, $\upsa(\zeta)$, $\upsb(\zeta)$, $\Lama(\zeta)$ and $\Lamb(\zeta)$ are shown in Fig.~\ref{fig:3jellium-HDL-SSF}, highlighting the Hoffmann minimum \cite{Hoffman92} in $\upsa(\zeta)$ near $\zeta=0.9956$ and revealing a similar minimum in $\Lama(\zeta)$ near $\zeta= 0.9960$.  It appears that such minima are ubiquitous in RPA coefficients.

%%% FIGURE 1 %%%

The data in Table \ref{tab:HDL} yield the exact values
\begin{subequations} 
\begin{align}
	\lam_1(0)	& = \frac{\alpha}{4\pi^3} \left(\frac{7\pi^2}{6} - 12 \ln 2 - 1\right)						\notag			\\
			& = 0.009\,229\ldots,																\label{eq:lam1-0}	\\
	\lam_1(1)	& = 2^{-4/3} \frac{\alpha}{4\pi^3} \left( \frac{13\pi^2}{12} -12\ln 2 +\frac{1}{2} \right)	\notag			\\
			& = 0.004\,792\ldots,																\label{eq:lam1-1}
\end{align}
\end{subequations} 
and it is revealing to compare these with recent numerical calculations.  
The estimate $\lam_1(0) \approx 0.0092292$ by Sun \textit{et al.} \cite{Sun10} agrees perfectly with Eq.~\eqref{eq:lam1-0} but their estimate $\lam_1(1) \approx 0.003125$ is strikingly different from Eq.~\eqref{eq:lam1-1}.  
The error arises from the non-commutivity of the $\zeta\to1$ limit and the $u$ integration, which is due to the non-uniform convergence of $\mathcal{R}_1^{\text{a}}(u,\zeta)$.

Based on the work of Carr and Maradundin, \cite{Carr64} Endo et al. \cite{Endo99} have been able to obtain a numerical value
\begin{equation}
\label{eq:eps1-3D}
	\eps_1(0) = -0.010
\end{equation}
for the paramagnetic limit of the term proportional to $\rs$.  However, nothing is known about the spin-scaling function and the ferromagnetic value for this coefficient.
Calculations by one of the present authors suggests that the value \eqref{eq:eps1-3D} is probably not accurate, \cite{Metal} \alert{mainly due to the large errors in the numerical integrations performed in Ref.~\onlinecite{Carr64}.}

%*************************
\subsubsection*{2-jellium}
%*************************
Gell-Mann--Brueckner resummation for 2-jellium yields \cite{Rajagopal77}
\begin{subequations} 
\begin{align}
\label{eq:lam0-0-2D}
	\lam_0(\zeta) 	
	& = 0,
	\\
\label{eq:lam1-0-2D}
	\lam_1(\zeta) 	
	& = - \frac{1}{12\sqrt{2}\pi} \int_{-\infty}^{\infty} \left[R\left(\frac{u}{\ku}\right)+R\left(\frac{u}{\kd}\right)\right]^3 du,
\end{align}
\end{subequations} 
where
\begin{equation}
\label{eq:Ru-2D}
	R(u) = 1 - \frac{1}{\sqrt{1+1/u^2}}.
\end{equation}
After an unsuccessful attempt by Zia, \cite{Zia73} the correct values of the coefficients $\lam_1(0)$ and $\lam_1(1)$ were found by Rajagopal and Kimball \cite{Rajagopal77} to be
\begin{equation}
	\lam_1(0) 
	= -\sqrt{2}\left(\frac{10}{3\pi}-1\right)
	= -0.086\,314\ldots,
\end{equation}
and \cite{Misawa65}
\begin{equation}
\label{eq:lam1-1-2D}
	\lam_1(1) 	
	= \frac{\sqrt{2}}{8} \lam_1(0)
	= - \frac{1}{4}\left(\frac{10}{3\pi}-1\right)
	= -0.015\,258\ldots.
\end{equation}
Thirty years later, Chesi and Giuliani found \cite{Chesi07} the spin-scaling function
\begin{equation}
\label{eq:L1-2D}
	\Lambda_1(\zeta) = \frac{\lam_1(\zeta)}{\lam_1(0)}
	= \frac{1}{8} \left[\ku+\kd+3 \frac{F\left(\ku,\kd\right)+F\left(\kd,\ku\right)}{10-3\pi}\right]
\end{equation}
where
\begin{equation}
	F(x,y) = 4(x+y) - \pi x 
%	\\
	- 4 x E\left(1-\frac{y^2}{x^2}\right) 
	+ 2x^2 \frac{\arccos \frac{y}{x}}{\sqrt{x^2-y^2}},	\label{eq:F}	
\end{equation}
and $E(x)$ is the complete elliptic integral of the second kind. \cite{NISTbook}

\begin{figure}
\includegraphics[width=0.45\textwidth]{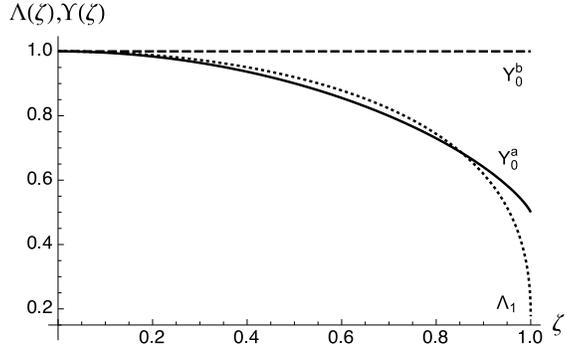}
\caption{
\label{fig:2jellium-HDL-SSF}
Spin-scaling functions of 2-jellium as functions of $\zeta$.}
\end{figure}

As in 3-jellium, the constant term $\eps_0(\zeta)$ can be decomposed into a direct contribution $\eps_{0}^{\text{a}}(\zeta)$ and a $\zeta$-independent exchange contribution $\eps_0^{\text{b}}$ 
\begin{equation}
\label{eq:eps0-2D}
	\eps_0(\zeta) = \eps_{0}^{\text{a}}(\zeta) + \eps_{0}^{\text{b}}.
\end{equation}
Following Onsager's work on the 3D case, \cite{Onsager66} Isihara and Ioriatti showed \cite{Isihara80} that
\begin{equation}
\label{eq:eps0-x-2D}
	\eps_{0}^\text{b} = 
	\beta(2) - \frac{8}{\pi^2}\beta(4) = +0.114\,357\ldots,
\end{equation}
where $G=\beta(2)$ is the Catalan's constant and $\beta$ is the Dirichlet beta function. \cite{NISTbook}  Recently, we have found closed-form expressions for the direct part $\eps_{0}^{\text{a}}(\zeta)$. \cite{2DEG11}
The paramagnetic and ferromagnetic limits are
\begin{subequations} 
\begin{align}
\label{eq:eps0-d-0-2D}
	\eps_{0}^{\text{a}}(0) & = \ln 2 - 1 = -0.306\,853\ldots,
	\\
\label{eq:eps0-d-1-2D}
	\eps_{0}^{\text{a}}(1) 
	& = \frac{1}{2} \eps_{0}^{\text{a}}(0)
	= \frac{\ln 2-1}{2} = -0.153\,426\ldots,
\end{align}
\end{subequations} 
and the spin-scaling functions are
\begin{equation}
\begin{split}
\label{eq:E0a-2D}
	\Upsilon_0^\text{a}(\zeta) 
	& = \frac{1}{2} + \frac{1-\zeta}{4(\ln 2 - 1)} 
	\Bigg[
	2\ln 2 - 1 
	\\
	& - \sqrt{\frac{1+\zeta}{1-\zeta}} + \frac{1+\zeta}{1-\zeta}  \ln\left(1 + \sqrt{\frac{1-\zeta}{1+\zeta}}\right) 
	\\
	& - \ln\left(1 + \sqrt{\frac{1+\zeta}{1-\zeta}}\right)
	\Bigg].
\end{split}
\end{equation}
and $\Upsilon_0^\text{b}(\zeta) = 1$.  The spin-scaling functions of 2-jellium are plotted in Fig.~\ref{fig:2jellium-HDL-SSF}.
To the best of our knowledge, the term proportional to $\rs$ in the high-density expansion of the correlation energy \eqref{eq:EcHDL} is unknown for 2-jellium.

%%% FIGURE 2 %%%

%*************************
\subsubsection*{1-jellium}
%*************************
Again, due to the strong divergence of the Coulomb operator in 1D, 1-jellium is peculiar and one has to take special care. \cite{1DChem15}
More details can be found in Ref.~\onlinecite{1DEG13}.
The leading term of the high-density correlation energy in 1-jellium has be found to be \cite{1DEG13}
\begin{equation}
	\eps_0 = -\frac{\pi^2}{360} = -0.027\,416\ldots,
\end{equation}
and third-order perturbation theory gives \cite{Carr64, 1DEG13}
\begin{equation}
\label{eq:eps1-1D}
	\eps_1 = +0.008\,446.
\end{equation}
We note that 1-jellium is one of the few systems where the $\rs$ coefficient of the high-density expansion is known accurately. \cite{Endo99, Sun10}
Unlike 2- and 3-jellium, the expansion \eqref{eq:EcHDL} does not contain any logarithm term up to first order in $\rs$, i.e. $\lambda_0 = \lambda_1 = 0$.
The high-density expansion of the correlation of 1-jellium is
\begin{equation}
\label{eq:1jellium-Ec-HDL}
	\ec(\rs) = - \frac{\pi^2}{360} + 0.008\,446\,\rs + \ldots.
\end{equation}

%----------------------------------------------------------------
\section{
\label{sec:LDL}
The low-density regime}
%----------------------------------------------------------------
In the low-density \alert{(or strongly-correlated)} regime, the potential energy dominates over the kinetic energy and the electrons localize onto lattice points that minimize their (classical) Coulomb repulsion. \cite{SGF14, LowGlo15}  
These minimum-energy configurations are called Wigner crystals (WC). \cite{Wigner34} 
In this regime, strong-coupling methods \cite{TEOAS09} can be used to show that the WC energy has the asymptotic expansion
\begin{equation}
\label{eq:eWC}
	\eWC(\rs) 
	\sim \sum_{j=0}^\infty \frac{\eta_j}{\rs^{j/2+1}}
	= \frac{\eta_0}{\rs} + \frac{\eta_1}{\rs^{3/2}} + \frac{\eta_2}{\rs^2} + \frac{\eta_3}{\rs^{5/2}} + \ldots.
\end{equation}
This equation is usually assumed to be strictly independent of the spin polarization. \cite{VignaleBook, AguileraNavarro85, Perdew92, Sun10}
\alert{The values of the low-density coefficients for $D$-jellium are reported in Table \ref{tab:LDL}}

%%% TABLE 3 %%%
\begin{table}
\caption{
\label{tab:LDL}
Energy coefficients of $D$-jellium at low density.
Note that $\gamma$ is the Euler-Mascheroni constant. \cite{NISTbook}}
\begin{ruledtabular}
\begin{tabular}{lcccc}
	Term			& 	Coeff.		&	\mc{1}{c}{$D = 3$}			&	\mc{1}{c}{$D = 2$}				&	\mc{1}{c}{$D = 1$}				\\
				& 				&	bcc lattice					&	$\triangle$ lattice				&	linear lattice			\\
\hline		
	$\rs^{-1}$		&	$\eta_0$		&	$-0.895\,930$	&	$-1.106\,103$		&	$(\gamma - \ln 2)/2$		\\
	$\rs^{-3/2}$	&	$\eta_1$		&	$1.325$	&	$0.795$			&	$0.359933$			\\
	$\rs^{-2}$		&	$\eta_2$		&	$-0.365$		&	unknown			&	unknown				\\
\end{tabular}
\end{ruledtabular}
\end{table}

%----------------------------------------------------------------
\subsection*{3-jellium}
%----------------------------------------------------------------
The leading term of the low-density expansion $\eta_0$ is the Madelung constant for the Wigner crystal. \cite{Fuchs35}
In 3D, Coldwell-Horsfall and Maradudin have studied several lattices: simple cubic (sc), face-centered cubic (fcc) and body-centered cubic (bcc).
Carr also mentions \cite{Carr61a} a calculation for the hexagonal closed-pack (hcp) by Kohn and Schechter. \cite{Kohn61}
The values of $\eta_0$ for these lattices are 
\begin{subequations}
\begin{align}
	\eta_0^\text{sc} & = -0.880\,059\ldots,
	\\
	\eta_0^\text{hcp} & = -0.895\,838\ldots,
	\\
	\eta_0^\text{fcc} & = -0.895\,877\ldots,
	\\
	\eta_0^\text{bcc} & = -0.895\,930\ldots.
	\label{eq:3jellium-eta0}
\end{align}
\end{subequations}
and reveal that, although all four lattices are energetically similar, the bcc lattice is the most stable.

For the bcc WC, Carr subsequently derived \cite{Carr61a} the harmonic zero-point energy coefficient 
\begin{equation}
\label{eq:3jellium-eta1}
	\eta_1 = 1.325,
\end{equation}
and the first anharmonic coefficient \cite{Carr61b} 
\begin{equation}
\label{eq:3jellium-eta2}
	\eta_2 = -0.365.
\end{equation}
Based on an interpolation, Carr and coworkers \cite{Carr61b} estimated the next term of the low-density asymptotic expansion to be $\eta_3 \approx -0.4$.

Combining Eqs.~\eqref{eq:3jellium-eta0}, \eqref{eq:3jellium-eta1} and \eqref{eq:3jellium-eta2} yields the low-density energy expansion of the 3D bcc WC
\begin{equation}
\label{eq:3jellium-Ec-LDL}
	\eWC(\rs) \sim  - \frac{0.895\,930}{\rs} + \frac{1.325}{\rs^{3/2}} - \frac{0.365}{\rs^2} + \ldots.
\end{equation}

%----------------------------------------------------------------
\subsection*{2-jellium}
%----------------------------------------------------------------
Following the same procedure as for 3-jellium, Bonsall and Maradundin \cite{Bonsall77} derived the leading term of the low-density energy expansion of the 2D WC for the square ($\square$) and triangular ($\triangle$) lattices:
\begin{subequations} 
\begin{align}
\begin{split}
	\eta_0^\square	& = - \frac{1}{\sqrt{\pi}} \left\{ 2 - {\sum_{\ell_1,\ell_2}}^\prime E_{-1/2}\left[ \pi (\ell_1^2 + \ell_2^2) \right] \right\}
	\\
					& = -1.100\,244\ldots,
\end{split}
\\
\begin{split}	
	\eta_0^\triangle 	& = - \frac{1}{\sqrt{\pi}} \left\{ 2 - {\sum_{\ell_1,\ell_2}}^\prime E_{-1/2}\left[ \frac{2\pi}{\sqrt{3}} (\ell_1^2 - \ell_1 \ell_2 + \ell_2^2) \right] \right\}
	\\
					& = -1.106\,103\ldots,
\end{split}
\end{align}
\end{subequations} 
where 
\begin{equation}
	E_{-1/2}(x) = \frac{1}{x} \left(\frac{\sqrt{\pi}}{2} \frac{\erfc(\sqrt{x})}{\sqrt{x}} + e^{-x}\right),
\end{equation}
erfc is the complementary error function \cite{NISTbook} and the prime excludes $(\ell_1,\ell_2)=(0,0)$ from the summation.
This shows that the triangular (hexagonal) lattice is more stable than the square one.

For the triangular lattice, Bonsall and Maradundin \cite{Bonsall77} also derived the harmonic coefficient
\begin{equation}
	\eta_1 = 0.795,
\end{equation}
but, to our knowledge, the first anharmonic coefficient is unknown.  This yields the 2D WC energy expression
\begin{equation}
\label{eq:2jellium-Ec-LDL}
	\eWC(\rs) \sim - \frac{1.106\,103}{\rs} + \frac{0.795}{\rs^{3/2}} + \ldots.
\end{equation}

%----------------------------------------------------------------
\subsection*{1-jellium}
%----------------------------------------------------------------
The first two coefficients of the low-density energy expansion of 1-jellium can be found in Fogler's work. \cite{Fogler05a}
The present authors have also given an alternative, simpler derivation using uniformly spaced electrons on a ring. \cite{Ringium13, Wirium14}
Both constructions lead to
\begin{subequations} 
\begin{align}
	\eta_0 & =  \frac{\gamma - \ln 2}{2} = -0.057\,966\ldots,
	\\
	\eta_1 
		& = \frac{1}{4\pi} \int_0^\pi \sqrt{2 \Li_3(1) - \Li_3(e^{i \theta}) - \Li_3(e^{-i \theta})} d\theta \notag
		\\
		& = +0.359\,933\ldots.
\end{align}
\end{subequations}
where $\Li_3$ is the trilogarithm function \cite{NISTbook} and the energy expansion
\begin{equation}
\label{eq:1jellium-Ec-LDL}
	\eWC(\rs) \sim \frac{\gamma - \ln 2}{2\rs} + \frac{0.359\,933}{\rs^{3/2}} + \ldots.
\end{equation}

%----------------------------------------------------------------
\section{
\label{sec:IDR}
The intermediate-density regime}
%----------------------------------------------------------------

%----------------------------------------------------------------
\subsection{
\label{sec:QMC}
Quantum Monte Carlo}
%----------------------------------------------------------------
Whereas it is possible to obtain information on the high- and low-density limits using perturbation theory, this approach struggles in the intermediate-density regime because of the lack of a suitable reference.  As a result, Quantum Monte Carlo (QMC) techniques \cite{Foulkes01, Kolorenc11} and, in particular, diffusion Monte Carlo (DMC) calculations have been valuable in this density range.  The first QMC calculations on 2- and 3-jellium were reported in 1978 by Ceperley. \cite{Ceperley78} \alert{Although QMC calculations have limitations (finite-size effect, \cite{Fraser96, Lin01, Kwee08, Drummond08, Ma11} fixed-node error, \cite{Ceperley91, Bressanini01, Bajdich05, Bressanini05a, Bressanini05b, Mitas06, Scott07, Bressanini08, Mitas08, Bressanini12, Rasch12, Kulahlioglu14, Rasch14, Nodes15} etc), these paved} the way for much subsequent research on the UEG and, indirectly, on the development of DFT. \cite{ParrBook}

%*************************
\subsubsection*{
\label{sec:QMC-3jellium}
3-jellium}
%*************************
Two years after Ceperley's seminal paper, \cite{Ceperley78} Ceperley and Alder published QMC results \cite{Ceperley80} that were subsequently used by various authors \cite{Vosko80, Perdew81, Perdew92} to construct UEG correlation functionals.  In their paper, Ceperley and Alder published released-node DMC results for the paramagnetic and ferromagnetic FF as well as the Bose fluid and bcc crystal.  Using these data, they proposed the first \alert{complete} phase diagram of 3-jellium and, despite its being based on a Bose bcc crystal, it is \alert{more than} qualitatively correct, as we will show later.  In particular, they found that 3-jellium has two phase transitions: a polarization transition (from paramagnetic to ferromagnetic fluid) at $\rs = 75 \pm 5$ and a ferromagnetic fluid-to-crystal transition at $\rs= 100 \pm 20$.

In the 1990's, Ortiz and coworkers extended Ceperley's study to partially-polarized fluid. \cite{Ortiz94, Ortiz97, Ortiz99}
They discovered a continuous transition from the paramagnetic to the ferromagnetic state in the range $20 \pm 5 \le \rs \le 40 \pm 5$ and they also predicted a much lower crystallization density ($\rs = 65\pm10$) than Ceperley and Alder.

Using more accurate trial wave function (with backflow) \cite{Kwon98} and twist-averaged boundary conditions \cite{Lin01} (to minimize finite-size effects), Zong et al. \cite{Zong02} re-evaluated the energy of the paramagnetic, ferromagnetic and partially-polarized fluid at relatively low density ($40 \le \rs \le 100$).
They found a second-order transition to a ferromagnetic phase at $\rs = 50 \pm 2$.
According to their results, the ferromagnetic fluid becomes more stable than the paramagnetic one at $\rs \approx 80$.

To complete the picture, Drummond et al. \cite{Drummond04} reported an exhaustive and meticulous study of the 3D WC over the range $100 \le \rs \le 150$.
They concluded that 3-jellium undergoes a transition from a ferromagnetic fluid to a bcc WC at $\rs = 106 \pm 1$, confirming the early prediction of Ceperley and Alder. \cite{Ceperley80} 
The discrepancy between the crystallization density found by Ortiz et al. \cite{Ortiz99} and the one determined by Drummond et al. \cite{Drummond04} is unclear.\alert{\footnote{\alert{The difference between the crystallisation densities reported in Ref.~\onlinecite{Ortiz99} and Ref.~\onlinecite{Ceperley80} is less than two error bars, whereas the crystallisation density difference between Ref.~\onlinecite{Ortiz99} and Ref.~\onlinecite{Drummond04} is of greater significance.}}}
The latter authors have also investigated the possibility of the existence of an antiferromagnetic WC phase but, sadly, they concluded that the energy difference between the ferromagnetic and antiferromagnetic crystals was too small to resolve in their DMC calculations.
More recently, Spink et al. \cite{Spink13} have also reported very accurate DMC energies for the partially-polarized fluid phase at moderate density ($0.2 \le \rs \le 20$).

\begin{figure}
\includegraphics[width=0.45\textwidth]{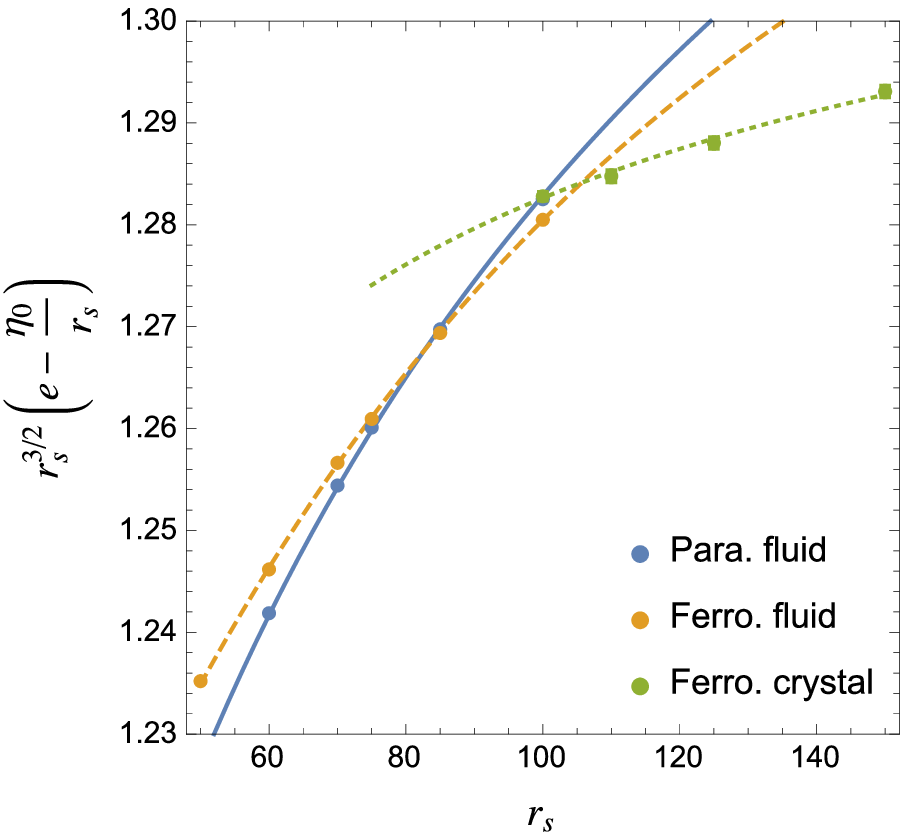}
\caption{
DMC phase diagram of 3-jellium.
\label{fig:3jellium-DMC-PD}
}
\end{figure}

The DMC energies of 3-jellium (for the FF and WC phases) have been gathered in Table \ref{tab:3jellium-QMC} for various $\rs$ and $\zeta$ values.
Combining the DMC results of Zong et al. \cite{Zong02} and Drummond et al., \cite{Drummond04} we have represented the phase diagram of 3-jellium in Fig.~\ref{fig:3jellium-DMC-PD}.
The correlation energy of the paramagnetic and ferromagnetic fluids is fitted using the parametrization proposed by Ceperley \cite{Ceperley78}
\begin{equation}
\label{eq:fit-fluid-C78}
	\ec(\rs) = \frac{a_0}{1 + a_1 \sqrt{\rs} + a_2 \rs},
\end{equation}
where $a_0$, $a_1$ and $a_2$ are fitting parameters.
For the ferromagnetic fluid, we have used the values of $a_0$, $a_1$ and $a_2$ given in Ref.~\onlinecite{Drummond04}.
These values have been obtained by fitting the ferromagnetic results of Zong et al. \cite{Zong02}
For the paramagnetic state, we have fitted the paramagnetic results of Ref.~\onlinecite{Zong02}, and found the values given in Table \ref{tab:3jellium-fit}.

To parametrize the WC energy data, Drummond et al. \cite{Drummond04} used another expression proposed by Ceperley \cite{Ceperley78}
\begin{equation}
\label{eq:fit-crystal-C78}
	\eWC(\rs) = \frac{b_0}{\rs} +  \frac{b_1}{\rs^{3/2}} +  \frac{b_2}{\rs^2}.
\end{equation}
The first coefficient $b_0$ is taken to be equal to the low-density limit expansion $\eta_0$ (see Sec.~\ref{sec:LDL}), while $b_1$ and $b_2$  are obtained by fitting the DMC results of Ref.~\onlinecite{Drummond04}.

%%% TABLE 4 %%%
\begin{table}
\caption{
Values of the coefficients $a_0$, $a_1$ and $a_2$ in Eq.~\eqref{eq:fit-fluid-C78} and $b_0$, $b_1$ and $b_2$ in Eq.~\eqref{eq:fit-crystal-C78} used to parametrize the energy of 3-jellium in the FF and WC phases.
\label{tab:3jellium-fit}}
\begin{ruledtabular}
\begin{tabular}{ccccc}
	\mc{3}{c}{Fermi fluid}												& 	\mc{2}{c}{Wigner crystal}		\\	
	\cline{1-3}												\cline{4-5}
	Coefficient	&	Para. 			&	Ferro. 			&	Coefficient	&	Ferro. 				\\
	\hline
	$a_0$		&	$-0.214488$		&	$-0.093 99$		&	$b_0$		&	$-0.89593$				\\
	$a_1$		&	$1.68634$		&	$1.5268$		&	$b_1$		&	$1.3379$				\\
	$a_2$		&	$0.490538$		&	$0.28882$		&	$b_2$		&	$-0.552 70$				\\
\end{tabular}
\end{ruledtabular}
\end{table}

%%% FIGURE 3 %%%

%%% TABLE 5 %%%
\begin{turnpage}
\begin{table*}
\caption{
\label{tab:3jellium-QMC}
DMC energy of 3-jellium at various $\rs$ for the FF and WC phases.
For the FF, the data from $\rs = 0.5$ to $20$ are taken from Ref.~\onlinecite{Spink13}, and the data from $\rs = 40$ to $100$ are taken from Ref.~\onlinecite{Zong02}.
The data for the ferromagnetic WC are taken from Ref.~\onlinecite{Drummond04}.  The statistical error is reported in parenthesis.}
\begin{ruledtabular}
\begin{tabular}{lllllllll}
$\rs$					&	Para. fluid				&	\mc{5}{c}{Partially spin-polarized fluid}																&	Ferro. fluid	&	Ferro. crystal									\\
					\cline{2-2}				\cline{3-7}																							\cline{8-8}			\cline{9-9}
					&	$\zeta=0$			&	$\zeta=0.185	$	&	$\zeta=0.333$		&	$\zeta=0.519$	&	$\zeta=0.667	$	&	$\zeta=0.852	$	&	$\zeta=1$		&	$\zeta=1$								\\
\hline
	0.5				&	$\phantom{-}3.430\,11(4)$		&	\cdash				&	$\phantom{-}3.692\,87(6)$		&	\cdash				&	$\phantom{-}4.441\,64(6)$		&	\cdash				&	$\phantom{-}5.824\,98(2)$		&	\cdash				\\
	1				&	$\phantom{-}0.587\,80(1)$		&	\cdash				&	$\phantom{-}0.649\,19(2)$		&	\cdash				&	$\phantom{-}0.823\,94(4)$		&	\cdash				&	$\phantom{-}1.146\,34(2)$		&	\cdash				\\
	2				&	$\phantom{-}0.002\,380(5)$		&	\cdash				&	$\phantom{-}0.016\,027(6)$		&	\cdash				&	$\phantom{-}0.054\,75(2)$		&	\cdash				&	$\phantom{-}0.126\,29(3)$		&	\cdash				\\
	3				&	$-0.067\,075(4)$		&	\cdash				&	$-0.061\,604(5)$		&	\cdash				&	$-0.046\,08(2)$		&	\cdash				&	$-0.017\,278(4)$		&	\cdash				\\
	5				&	$-0.075\,881(1)$		&	\cdash				&	$-0.074\,208(4)$		&	\cdash				&	$-0.069\,548(4)$		&	\cdash				&	$-0.060\,717(5)$		&	\cdash				\\
	10				&	$-0.053\,511\,6(5)$	&	\cdash				&	$-0.053\,214(2)$		&	\cdash				&	$-0.052\,375(2)$		&	\cdash				&	$-0.050\,733\,7(5)$	&	\cdash				\\
	20				&	$-0.031\,768\,6(5)$	&	\cdash				&	$-0.031\,715\,6(7)$	&	\cdash				&	$-0.031\,594\,0(7)$	&	\cdash				&	$-0.031\,316\,0(4)$	&	\cdash				\\
	40				&	$-0.017\,618\,7(3)$	&	\cdash				&	$-0.017\,6165(3)$		&	\cdash				&	$-0.017\,602\,7(3)$	&	\cdash				&	$-0.017\,567\,4(4)$	&	\cdash				\\
	50				&	$-0.014\,449\,5(3)$	&	$-0.014\,449\,5(3)$	&	$-0.014\,449\,8(3)$	&	$-0.014\,447\,3(4)$	&	$-0.014\,444\,2(3)$	&	$-0.014\,437\,7(4)$	&	$-0.014\,424\,9(4)$	&	\cdash				\\
	60				&	$-0.012\,260\,1(2)$	&	$-0.012\,259\,3(3)$	&	$-0.012\,260\,2(2)$	&	$-0.012\,259\,8(3)$	&	$-0.012\,258\,7(2)$	&	$-0.012\,255\,9(2)$	&	$-0.012\,250\,8(2)$	&	\cdash				\\
	70				&	$-0.010\,657\,2(2)$	&	$-0.010\,656\,9(2)$	&	$-0.010\,658\,1(2)$	&	$-0.010\,658\,6(3)$	&	$-0.010\,658\,0(2)$	&	$-0.010\,656\,7(2)$	&	$-0.010\,653\,3(2)$	&	\cdash				\\
	75				&	$-0.010\,005\,7(2)$	&	$-0.010\,006\,0(2)$	&	$-0.010\,006\,9(2)$	&	\cdash				&	$-0.010\,007\,2(2)$	&	\cdash				&	$-0.010\,004\,4(2)$	&	\cdash				\\
	85				&	$-0.008\,920\,1(2)$	&	\cdash				&	$-0.008\,9208(2)$		&	\cdash				&	$-0.008\,921\,5(2)$	&	\cdash				&	$-0.008\,920\,6(2)$	&	\cdash				\\
	100				&	$-0.007\,676\,8(2)$	&	\cdash				&	$-0.007\,677(2)$		&	\cdash				&	$-0.007\,678\,2(1)$	&	\cdash				&	$-0.007\,678\,8(1)$	&	$-0.007\,676\,5(4)$	\\
	110				&	\cdash				&	\cdash				&	\cdash				&	\cdash				&	\cdash				&	\cdash				&	\cdash				&	$-0.007\,031\,2(5)$	\\
	125				&	\cdash				&	\cdash				&	\cdash				&	\cdash				&	\cdash				&	\cdash				&	\cdash				&	$-0.006\,245\,8(4)$	\\
	150				&	\cdash				&	\cdash				&	\cdash				&	\cdash				&	\cdash				&	\cdash				&	\cdash				&	$-0.005\,269\,0(3)$	\\
\end{tabular}
\end{ruledtabular}
\end{table*}
\end{turnpage}

%*************************
\subsubsection*{
\label{sec:QMC-2jellium}
2-jellium}
%*************************
The first exhaustive study of 2-jellium at the DMC level was published in 1989 by Tanatar and Ceperley. \cite{Tanatar89}
In their study, the authors investigate the paramagnetic and ferromagnetic fluid phases, as well as the ferromagnetic WC with hexagonal symmetry (triangular lattice).
They discovered a Wigner crystallization at $\rs = 37 \pm 5$ and they found that, although they are very close in energy, the paramagnetic fluid is always more stable than the ferromagnetic one.
Although the Tanatar-Ceperley energies are systematically too low, as noted by Kwon, Ceperley and Martin, \cite{Kwon93} their phase diagram is qualitatively correct. 

A few years later, Rapisarda and Senatore \cite{Rapisarda96} revisited the phase diagram of 2-jellium.
They found a region of stability for the ferromagnetic fluid with a polarization transition at $\rs = 20 \pm 2$ and observed a ferromagnetic fluid-to-crystal transition at $\rs = 34 \pm 4$.
This putative region of stability for the ferromagnetic fluid was also observed by Attaccalite et al. \cite{Attaccalite02, Attaccalite03, GoriGiorgi03} who obtained a similar phase diagram with a polarization transition at $\rs \approx 26$ and a crystallization at $\rs \approx 35$. 
An important contribution of Ref.~\onlinecite{Attaccalite02} was to show that, in contrast to 3-jellium, the partially-polarized FF is never a stable phase of 2-jellium.

More recently, and in contrast to earlier QMC studies, Drummond and Needs \cite{Drummond09} obtained statistical errors sufficiently small to resolve the energy difference between the ferromagnetic and paramagnetic fluids.
Interestingly, instead of observing a transition from the ferromagnetic fluid to the ferromagnetic crystal, they discovered a transition from the paramagnetic fluid to an antiferromagnetic crystal around $\rs = 31 \pm 1$.
Moreover, they also showed that the ferromagnetic fluid is never more stable than the paramagnetic one, and that it is unlikely that a region of stability exists for a partially spin-polarized fluid.  This agrees with the earlier work of Attaccalite et al. \cite{Attaccalite02}
However, they did find a transition from the antiferromagnetic to the ferromagnetic WC at $\rs = 38 \pm 5$.

Some authors have investigated the possibility of the existence of a ``hybrid phase'' in the vicinity of the transition density from ferromagnetic fluid to ferromagnetic WC. \cite{Spivak04, Falakshahi05, Waintal06, Clark09, Drummond09}  According to Falakshahi and Waintal, \cite{Falakshahi05, Waintal06} the hybrid phase has the same symmetry as the WC but has partially delocalized orbitals.
However, its existence is still under debate. \cite{Drummond09}

\begin{figure}
\includegraphics[width=0.45\textwidth]{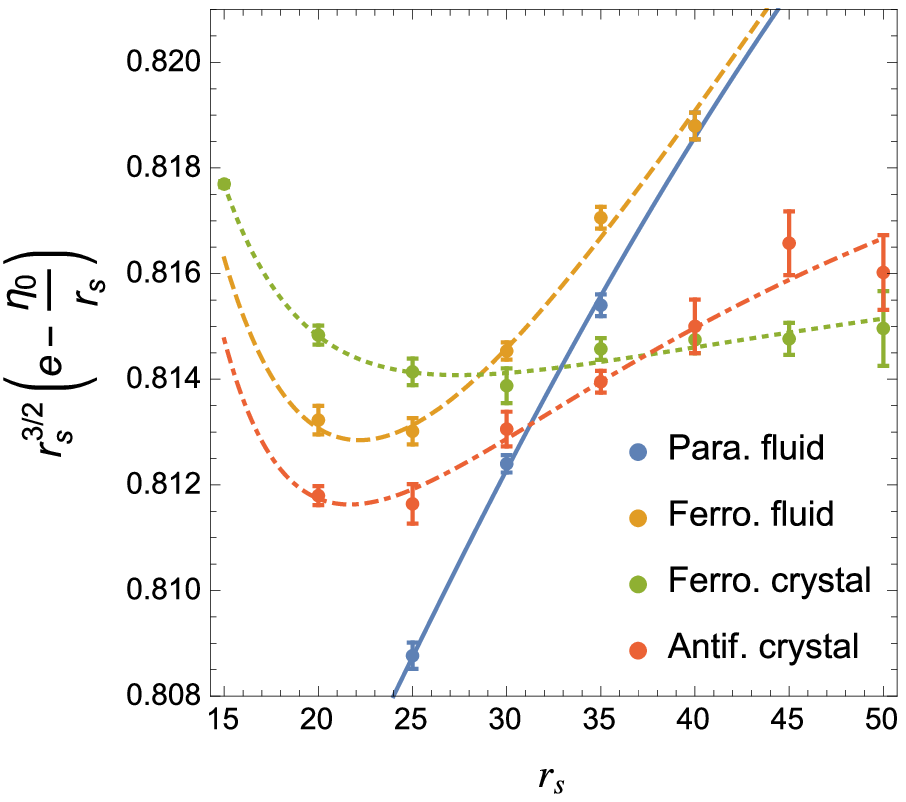}
\caption{
DMC phase diagram of 2-jellium.
\label{fig:2jellium-DMC-PD}
}
\end{figure}

The DMC energies of 2-jellium (for the fluid and crystal phases) have been gathered in Table \ref{tab:2jellium-QMC} for various $\rs$.
Based on the data of Ref.~\onlinecite{Drummond09}, we have constructed the phase diagram of 2-jellium in Fig.~\ref{fig:2jellium-DMC-PD}.
The fluid energy data are fitted using the parametrization proposed by Rapisarda and Senatore: \cite{Rapisarda96}
\begin{equation}
\label{eq:fit-fluid-RS96}
\begin{split}
	\ec(\rs)
				& = a_0 \left\{ 1 + A \rs \left[ B \ln \frac{\sqrt{\rs} + a_1}{\sqrt{\rs}} \right. \right. 
				\\
				& \qquad \qquad + \frac{C}{2} \ln \frac{\rs + 2 a_2 \sqrt{\rs} + a_3}{\rs} 
				\\
				& \qquad \qquad  \left. \left. + D \left( \arctan \frac{\sqrt{\rs} + a_2}{\sqrt{a_3 - a_2^2}} - \frac{\pi}{2} \right) \right] \right\},
\end{split}
\end{equation}
where 
\begin{subequations} 
\begin{align}
	A & = \frac{2(a_1+2a_2)}{2a_1a_2-a_3-a_1^2},
	\qquad  
	B = \frac{1}{a_1} - \frac{1}{a_1+2a_2},
	\\ 
	C & = \frac{a_1}{a_3} - \frac{2a_2}{a_3} + \frac{1}{a_1+2a_2},
	\qquad 
	D = \frac{F-a_2 C}{\sqrt{a_3-a_2^2}},
	\\ 
	F & = 1+(2a_2-a_1) \left( \frac{1}{a_1+2a_2}-\frac{2a_2}{a_3} \right).
\end{align}
\end{subequations} 

To parametrize the WC energies, Drummond and Needs \cite{Drummond09} used the expression proposed by Ceperley \cite{Ceperley78}
\begin{equation}
\label{eq:fit-crystal-RS96}
	\eWC(\rs) = \frac{b_0}{\rs} +  \frac{b_1}{\rs^{3/2}} +  \frac{b_2}{\rs^2} +  \frac{b_3}{\rs^{5/2}}+  \frac{b_4}{\rs^3}.
\end{equation}
The first two coefficients $b_0$ and $b_1$ are taken to be equal to the low-density limit expansion $\eta_0$ and $\eta_1$ (see Sec.~\ref{sec:LDL}), and the others are found by fitting to their DMC results.
\alert{The values of the fitting coefficients for 2-jellium are given in Table \ref{tab:2jellium-fit}.}

\begin{table*}
\caption{
Values of the coefficients $a_0$, $a_1$, $a_2$ and $a_3$ in Eq.~\eqref{eq:fit-fluid-RS96} and $b_0$, $b_1$, $b_2$ and $b_3$ and $b_4$ in Eq.~\eqref{eq:fit-crystal-RS96} used to parameterize the energy of 2-jellium in the FF and WC phases.
\label{tab:2jellium-fit}}
\begin{ruledtabular}
\begin{tabular}{cccccc}
	\mc{3}{c}{Fermi fluid}												& 	\mc{3}{c}{Wigner crystal}							\\	
	\cline{1-3}												\cline{4-6}
	Coefficient	&	\mc{2}{c}{Value}							&	Coefficient	&	\mc{2}{c}{Value}								\\
				\cline{2-3}													\cline{5-6}
				&	Para. fluid			&	Ferro. fluid		&				&	Ferro. crystal			&	Antif. crystal			\\
	\hline
	$a_0$		&	$-0.186\,305\,2$		&	$-0.290\,910\,2$	&	$b_0$		&	$-1.106\,103$		&	$-1.106\,103$		\\
	$a_1$		&	$6.821\,839$			&	$-0.624\,383\,6$	&	$b_1$		&	$0.814$				&	$0.814$				\\
	$a_2$		&	$0.155\,226$			&	$1.656\,628$		&	$b_2$		&	$0.113\,743$			&	$ 0.266\,297\,7$		\\
	$a_3$		&	$3.423\,013$			&	$3.791\,685$		&	$b_3$		&	$-1.184\,994$		&	$ -2.632\,86$		\\
				&						&					&	$b_4$		&	$3.097\,610$			&	$6.246\,358 $		\\
\end{tabular}
\end{ruledtabular}
\end{table*}

%%% FIGURE 4 %%%

%%% TABLE 7 %%%
\begin{table}
\caption{
\label{tab:2jellium-QMC}
DMC energy of 2-jellium at various $\rs$ for the FF and WC phases.
Data from $\rs = 1$ to $10$ are taken from Ref.~\onlinecite{Kwon93} for the paramagnetic fluid. 
Data from $\rs = 5$ to $10$ are taken from Ref.~\onlinecite{Rapisarda96} for the ferromagnetic fluid. 
Data from $\rs = 15$ to $50$ are taken from Ref.~\onlinecite{Drummond09}.
The statistical error is reported in parenthesis.}
\begin{ruledtabular}
\begin{tabular}{lllll}
	$\rs$		&	Para. fluid		&	Ferro. fluid			&	Antif. cystal		&	Ferro. crystal		\\
				\cline{2-2}				\cline{3-3}				\cline{4-4}				\cline{5-5}
				&	\mc{1}{c}{$\zeta=0$}			&	\mc{1}{c}{$\zeta=1$}		&	\mc{1}{c}{$\zeta=0$}		&	\mc{1}{c}{$\zeta=1$}		\\
\hline		
	1			&	$-0.209\,8(3)$			&	\cdash			&	\cdash			&	\cdash			\\
	5			&	$-0.149\,5(1)$			&	$-0.143\,3(1)$		&	\cdash			&	\cdash			\\
	10			&	$-0.085\,36(2)$			&	$-0.084\,48(4)$		&	\cdash			&	\cdash			\\
	15			&	\cdash				&	\cdash			&	\cdash			&	$-0.059\,665(1)$		\\
	20			&	$-0.046\,305(4)$		&	$-0.046\,213(3)$	&	$-0.046\,229(2)$	&	$-0.046\,195(2)$		\\
	25			&	$-0.037\,774(2)$		&	$-0.037\,740(2)$	&	$-0.037\,751(3)$	&	$-0.037\,731(2)$		\\
	30			&	$-0.031\,926(1)$		&	$-0.031\,913(1)$	&	$-0.031\,922(2)$	&	$-0.031\,917(2)$		\\
	35			&	$-0.027\,665(1)$		&	$-0.027\,657(1)$	&	$-0.027\,672(1)$	&	$-0.027\,669(1)$		\\
	40			&	$-0.024\,416(1)$		&	$-0.024\,416(1)$	&	$-0.024\,431(2)$	&	$-0.024\,432(1)$		\\
	45			&	\cdash				&	\cdash			&	$-0.021\,875(2)$	&	$-0.021\,881(1)$		\\
	50			&	\cdash				&	\cdash			&	$-0.019\,814(2)$	&	$-0.019\,817(2)$		\\
\end{tabular}
\end{ruledtabular}
\end{table}

%*************************
\subsubsection*{1-jellium}
%*************************
Not surprisingly, there have been only a few QMC studies on 1-jellium.
Astrakharchik and Girardeau \cite{Astrakharchik11} have studied 1-jellium qualitatively from the high to the low density regimes.
Lee and Drummond \cite{Lee11a} have published accurate DMC data for the range $1 \le \rs \le 20$.
The present authors have published DMC data at higher and lower densities in order to parametrize a generalized version of the LDA. \cite{Ringium13, gLDA14, Wirium14}
The DMC data for 1-jellium are reported in Table \ref{tab:1jellium-QMC}.

%%% TABLE 8 %%%
\begin{table}
\caption{
\label{tab:1jellium-QMC}
DMC energy and reduced energy given by Eq.~\eqref{eq:Ec-Cios} for 1-jellium at various $\rs$.
The DMC data from $\rs = 1$ to $20$ are taken from Ref.~\onlinecite{Lee11a}.
The rest is taken from Refs. \onlinecite{1DEG13, Ringium13, Wirium14}.
The statistical error is reported in parenthesis.}
\begin{ruledtabular}
\begin{tabular}{lll}	
	$\rs$	&	DMC energy					&	$\eHF + e_\text{c}^{\text{LDA}}$	\\
	\hline
	0.2		&	$\phantom{-}13.100\,54(2)$			&	$\phantom{-}13.100\,53$			\\
	0.5		&	$\phantom{-}1.842\,923(2)$			&	$\phantom{-}1.842\,850$			\\
	1		&	$\phantom{-}0.154\,188\,6(2)$			&	$\phantom{-}0.154\,101\,4$		\\
	2		& 	$-0.206\,200\,84(7)$			&	$-0.206\,219\,38$		\\
	5		&	$-0.203\,932\,35(2)$			&	$-0.203\,843\,14$		\\
	10		&	$-0.142\,869\,097(9)$		&	$-0.142\,781\,622$		\\
	15		&	$-0.110\,466\,761(4)$		&	$-0.110\,400\,702$		\\
	20		&	$-0.090\,777\,768(2)$		&	$-0.090\,727\,757$		\\
	50		&	$-0.046\,144(1)$			&	$-0.046\,128$			\\
	100		&	$-0.026\,699(1)$			&	$-0.026\,694$			\\
\end{tabular}
\end{ruledtabular}
\end{table}

Using the ``robust'' interpolation proposed by Cioslowski \cite{Cioslowski12} and the high- and low-density expansions \eqref{eq:1jellium-Ec-HDL} and \eqref{eq:1jellium-Ec-LDL}, the correlation energy of 1-jellium calculated with the HF energy given by \eqref{eq:eHF-1D} can be approximated by
\begin{equation}
\label{eq:Ec-Cios}
	e_\text{c}^\text{LDA}(\rs) = t^2 \sum_{j=0}^{3} c_j t^j (1-t)^{3-j},
\end{equation}
with 
\begin{equation}
	t = \frac{\sqrt{1+4\,k\,\rs}-1}{2\,k\,\rs},
\end{equation}
and
\begin{subequations} 
\begin{align}
	c_0 & = k\,\eta_0,	
	&
	c_1 & = 4\,k\,\eta_0+k^{3/2}\eta_1,
	\\
	c_2 & = 5\,\eps_0+\eps_1/k,
	&
	c_3 & = \eps_1,
\end{align}
\end{subequations} 
where $k=0.414254$ is a scaling factor which is determined by a least-squares fit of the DMC data given in Refs.~\onlinecite{Lee11a} and \onlinecite{Ringium13}. 

\begin{figure}
\includegraphics[width=0.45\textwidth]{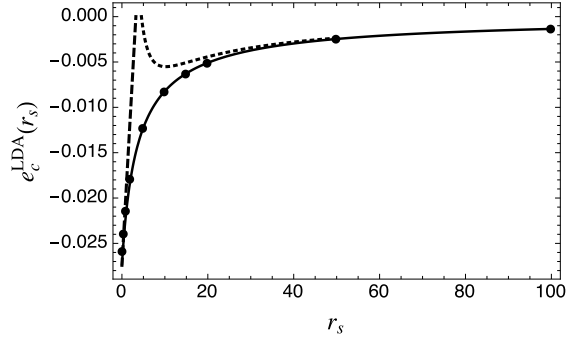}
\caption{
\label{fig:1jellium-fit}
$e_\text{c}^\text{LDA}(\rs)$ \alert{of 1-jellium} given by Eq.~\eqref{eq:Ec-Cios} as a function of $\rs$ (solid line).
DMC results from Table \ref{tab:1jellium-QMC} are shown by black dots. 
The small-$\rs$ expansion of Eq.~\eqref{eq:1jellium-Ec-HDL} (dashed line) and large-$\rs$ approximation of Eq.~\eqref{eq:1jellium-Ec-LDL} (dotted line) are also shown.}
\end{figure}

The results using the LDA correlation functional \eqref{eq:Ec-Cios} are compared to the DMC calculations of Refs.~\onlinecite{Lee11a} and \onlinecite{Ringium13}. 
The results are gathered in Table \ref{tab:1jellium-QMC} and depicted in Fig.~\ref{fig:1jellium-fit}.  
For $0.2 \le \rs \le 100$, the LDA and DMC correlation energies agree to within 0.1 millihartree, which is remarkable given the simplicity of the functional. 

%%% FIGURE 5 %%%

%----------------------------------------------------------------
\subsection{
\label{sec:SBHF}
Symmetry-broken Hartree-Fock}
%----------------------------------------------------------------
In the early 1960's, Overhauser \cite{Overhauser59, Overhauser62} showed that the HF energy \eqref{eq:eHF} for the paramagnetic FF can always be improved by following spin- and charge-density instabilities \cite{VignaleBook} to locate a symmetry-broken HF (SBHF) solution.  Recently, a computational ``proof'' has been given by Zhang and Ceperley \cite{Zhang08} who performed unrestricted HF (UHF) calculations on the paramagnetic state of finite-size 3D UEGs and discovered broken spin-symmetry solutions, even for high densities.
In 2D, this has been proven rigorously for the ferromagnetic state by Bernu et al. \cite{Bernu08}  The first phase diagrams based on UHF calculations for 2- and 3-jellium were performed by Trail et al. \cite{Trail03} who found lower energies for a crystal for $\rs > 1.44$ in 2D and $\rs > 4.5$ in 3D.
Curiously, as we will show below, the SBHF phase diagram is far richer than the near-exact DMC one presented in Sec.~\ref{sec:QMC}.

\begin{figure}
\includegraphics[width=0.45\textwidth]{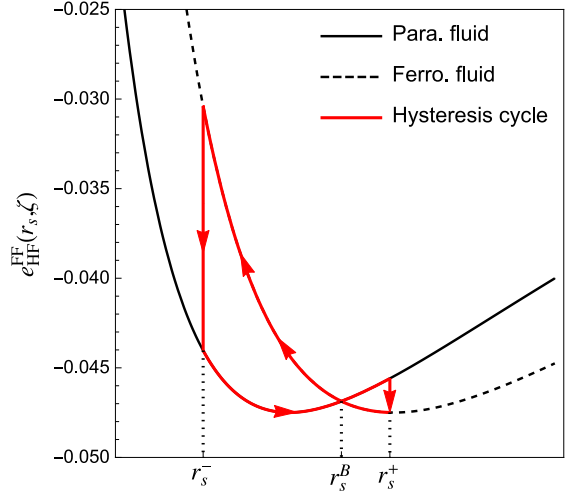}
\caption{
$\eHF(\rs,\zeta)$ as a function of $\rs$ for the paramagnetic and ferromagnetic fluid phases of 3-jellium (see Eq.~\eqref{eq:eHF}).
For $\rs  > \rs^\text{B}$, the ferromagnetic fluid becomes lower in energy than the paramagnetic fluid (Bloch transition).
For $\rs < \rs^-$, the ferromagnetic fluid becomes locally unstable towards depolarization, while for $\rs > \rs^+$, the paramagnetic fluid becomes locally unstable towards polarization.
The ``hysteresis loop'' is indicated in red.
\label{fig:HF-hysteresis}
}
\end{figure}

Before going further, it is interesting to investigate the HF expression of the FF given by \eqref{eq:eHF}, and study the phase diagram based on this simple expression \cite{VignaleBook} (see Fig.~\ref{fig:HF-hysteresis} for the example of 3-jellium).
It is easy to show that, for $0 < \rs < \rs^\text{B}$, the paramagnetic fluid is predicted to be lower in energy than the ferromagnetic fluid where
\begin{equation}
\label{eq:rsB}
	\rs^\text{B} = - \frac{2^{2/D}-1}{2^{1/D}-1} \frac{\eps_t}{\eps_x}
	=
	\begin{cases}
		2.011,	&	D = 2,	\\
		5.450,	&	D = 3,	\\		
	\end{cases}
\end{equation}
and $\eps_t$ and $\eps_x$ are given by Eqs.~\eqref{eq:epst} and \eqref{eq:epsx}, respectively.
This sudden paramagnetic-to-ferromagnetic transition is sometimes called a Bloch transition. \cite{Bloch29}
Expanding the HF expression of the paramagnetic state around $\zeta = 0$ yields
\begin{equation}
\begin{split}
	\eHF(\rs,\zeta) 
	& = \eHF(\rs,0) 
	\\
	& + \zeta^2 \left( \frac{D+2}{D^2} \frac{\eps_\text{t}}{\rs^2} + \frac{D+1}{2D^2} \frac{\eps_\text{x}}{\rs} \right) + O(\zeta^4).
\end{split}
\end{equation}
and reveals that this state is locally stable with respect to partial spin-polarization until 
\begin{equation}
\label{eq:rsplus}
	\rs^+ = - \frac{2(D+2)}{D+1} \frac{\eps_t}{\eps_x}
	=
	\begin{cases}
		2.221,	&	D = 2,	\\
		6.029,	&	D = 3.	\\		
	\end{cases}
\end{equation}
The fact that $\rs^+ > \rs^\text{B}$ implies that this state is locally stable with respect to partial spin-polarization and will not undergo a continuous phase transition to the ferromagnetic state, in contrast to the predictions of DMC calculations on 3-jellium, as discussed in Sec.~\ref{sec:QMC-3jellium}.

For $\rs > \rs^\text{B}$, the ferromagnetic state is lower in energy than the paramagnetic state.
However, a similar stability analysis yields
\begin{equation}
\begin{split}
	\eHF(\rs,\zeta) 
	& = \eHF(\rs,1) 
	\\
	& - (1-\zeta) \left(\frac{D+2}{2^{\frac{D-1}{D}} D} \frac{\eps_\text{t}}{\rs^2} +  \frac{D+1}{2^{\frac{D-1}{D}} D} \frac{\eps_\text{x}}{\rs} \right) 
	\\
	& + O\left((1-\zeta)^{\frac{D+1}{D}}\right),
\end{split}
\end{equation}
which shows that the ferromagnetic state is never a stationary minimum.
In fact, for $\rs < \rs^-$, where
\begin{equation}
\label{eq:rsminus}
	\rs^- = \frac{\rs^+}{2^{\frac{D-1}{D}}}
	=
	\begin{cases}
		1.571,	&	D = 2,	\\
		3.798,	&	D = 3,	\\		
	\end{cases}
\end{equation}
the ferromagnetic state is locally unstable and can undergo a continuous depolarization towards the paramagnetic state.
Taken together, these predictions imply the ``hysteresis loop'' shown in Fig.~\ref{fig:HF-hysteresis} for 3-jellium.

%%% FIGURE 6 %%%

%*************************
\subsubsection*{3-jellium}
\begin{figure}
\includegraphics[width=0.6\textwidth]{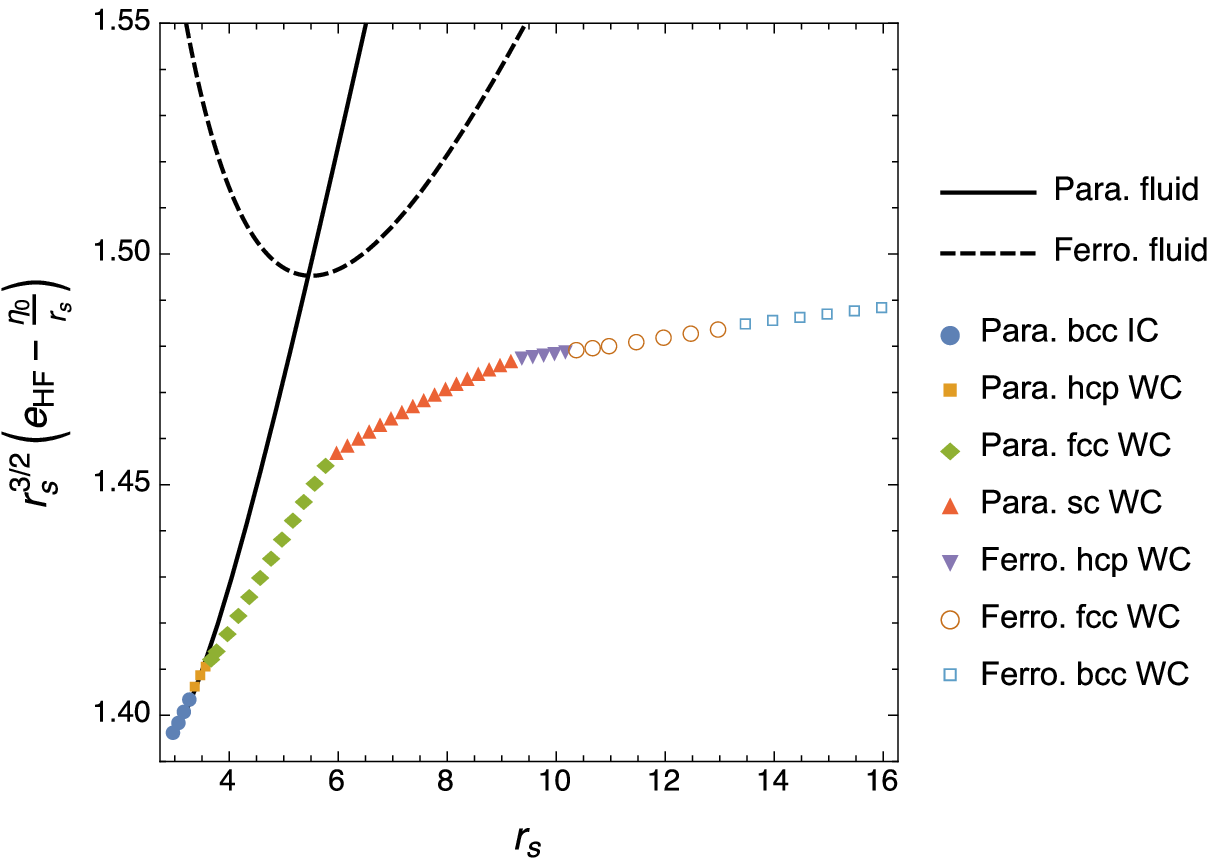}
\caption{
SBHF phase diagram of 3-jellium constructed with the data of Refs.~\onlinecite{Baguet13, Baguet14} (see Table \ref{tab:3jellium-SBHF}).
\label{fig:3jellium-HF-PD}
}
\end{figure}

%*************************
Baguet et al. \cite{Baguet13, Baguet14} have obtained what is thought to be the complete phase diagram of 3-jellium at the HF level.
The SBHF phase diagram of 3-jellium is represented in Fig.~\ref{fig:3jellium-HF-PD} using the data reported in Refs.~\onlinecite{Baguet13, Baguet14} (see Table \ref{tab:3jellium-SBHF}). 
In addition to the usual FF and WC phases, they have also considered incommensurate crystals (IC) with sc, fcc, bcc and hcp unit cells.
In an IC, the number of maxima of the charge density is higher than the number of electrons, having thus metallic character.
As one can see in  Fig.~\ref{fig:3jellium-HF-PD}, the phase diagram is complicated and, unfortunately, finite-size effects prevent a precise determination of the ground state for $\rs < 3$.
However, extending the analysis of Ref.~\onlinecite{Delyon08}, one can prove that the incommensurate phases are always energetically lower than the FF in the high-density limit.
\alert{This particular point has been recently discussed in Ref.~\onlinecite{Delyon15}.}

For $3 < \rs < 3.4$, the incommensurate metallic phase with a bcc lattice is found to be the lowest-energy state. 
For $\rs > 3.4$, the 3-jellium ground state is a paramagnetic WC with hcp ($3.4 < \rs < 3.7$), fcc  ($3.7 < \rs < 5.9$) and sc ($5.9 < \rs < 9.3$) lattices.
From any value of $\rs$ greater than $9.3$, the ground state is a ferromagnetic WC with hcp ($9.3 < \rs < 10.3$), fcc ($10.3 < \rs < 13$) and finally bcc ($\rs > 13$) lattices.
It is interesting to note that, compared to the DMC results from Sec.~\ref{sec:QMC-3jellium}, at the HF level, the Wigner crystallization happens at much higher densities, revealing a key deficiency of the HF theory.

%%% FIGURE 7 %%%

%%% TABLE 9 %%%
\begin{table*}
\caption{
\label{tab:3jellium-SBHF}
SBHF energy (in millihartree) of 3-jellium for various $\rs$ values.
The energy data are taken from the supplementary materials of Ref.~\onlinecite{Baguet13}.
The precision of the calculations is of the order $5 \times 10^{-3}$ millihartree.
}
\begin{ruledtabular}
\begin{tabular}{lcccc|lcccc}
$\rs$		&	Energy		&	Lattice	&	Phase		&	Polarization	&	$\rs$		&	Energy		&	Lattice	&	Phase		&	Polarization	\\
\hline		
 3.0		&	 $-29.954 $	&	bcc 		&	IC 	&	Para.	&	 7.6 		&	  $-47.804 $	&	sc 	&	WC	&	Para.	\\
 3.1		&	 $-32.826 $	&	bcc 		&	IC 	&	Para.	&	 7.8 		&	  $-47.403 $	&	sc 	&	WC	&	Para.	\\
 3.2		&	 $-35.289 $	&	bcc 		&	IC 	&	Para.	&	 8.0 		&	  $-46.992 $ 	&	sc 	&	WC	&	Para.	\\
 3.3 		&	 $-37.399 $	&	bcc		&	IC	&	Para.	&	 8.2 		&	  $-46.576 $ 	&	sc 	&	WC	&	Para.	\\
 3.4 		&	 $-39.287 $	&	hcp	 	&	WC 	&	Para.	&	 8.4 		&	  $-46.155 $ 	&	sc	&	WC	&	Para.	\\
 3.5 		&	 $-40.923 $	&	hcp 		&	WC 	&	Para.	&	 8.6 		&	  $-45.731 $	&	sc	&	WC	&	Para.	\\
 3.6 		&	 $-42.437 $	&	hcp 		&	WC	&	Para.	&	 8.8 		&	  $-45.307 $	&	sc	&	WC	&	Para.	\\
 3.7 		&	 $-43.727 $	&	fcc 		&	WC	&	Para.	&	 9.0		&	  $-44.883 $	&	sc 	&	WC	&	Para.	\\
 3.8 		&	 $-44.899 $	&	fcc 		&	WC	&	Para.	&	 9.2 		&	  $-44.461 $	&	sc	&	WC	&	Para.	\\
 4.0 		&	 $-46.775 $	&	fcc		&	WC	&	Para.	&	 9.4 		&	  $-44.050 $	&	hcp	&	WC	&	Ferro.	\\
 4.2 		&	 $-48.157 $ 	&	fcc		&	WC	&	Para.	&	 9.6 		&	  $-43.647 $ 	&	hcp 	&	WC	&	Ferro.	\\
 4.4 		&	 $-49.151 $ 	&	fcc 		&	WC	&	Para.	&	 9.8 		&	  $-43.245 $ 	&	hcp	&	WC  	&	Ferro.	\\
 4.6 		&	 $-49.841 $ 	&	fcc 		&	WC	&	Para.	&	 10.0 	&	  $-42.844 $ 	&	hcp 	&	WC  	&	Ferro.	\\ 
 4.8 		&	 $-50.292 $ 	&	fcc 		&	WC	&	Para.	&	 10.2 	&	  $-42.444 $ 	&	hcp 	&	WC	&	Ferro.	\\
 5.0 		&	 $-50.554 $ 	&	fcc 		&	WC	&	Para.	&	 10.4		&	  $-42.047 $ 	&	fcc 	&	WC	&	Ferro.	\\
 5.2 		&	 $-50.665 $ 	&	fcc 		&	WC	&	Para.	& 	 10.7 	& 	  $-41.461 $	& 	fcc 	&	WC	& 	Ferro.	\\
 5.4 		&	 $-50.656 $ 	&	fcc 		&	WC	&	Para.	&	 11.0 	&	  $-40.883 $	&	fcc 	&	WC	&	Ferro.	\\
 5.6 		&	 $-50.551 $ 	&	fcc 		&	WC	&	Para.	&	 11.5 	&	  $-39.936 $	&	fcc 	&	WC	&	Ferro.	\\
 5.8 		&	 $-50.368 $ 	&	fcc 		&	WC	&	Para.	&	 12.0 	&	  $-39.015 $	&	fcc	&	WC	&	Ferro.	\\
 6.0 		&	 $-50.192 $ 	&	sc 		&	WC	&	Para.	&	 12.5 	&	  $-38.126 $	&	fcc 	&	WC	&	Ferro.	\\
 6.2 		&	 $-50.031 $ 	&	sc 		&	WC	&	Para.	&	 13.0 	&	  $-37.267 $	&	fcc 	&	WC	&	Ferro.	\\
 6.4 		&	 $-49.813 $ 	&	sc 		&	WC	&	Para.	&	 13.5 	&	  $-36.441 $	&	bcc 	&	WC	&	Ferro.	\\
 6.6 		&	 $-49.550 $ 	&	sc 		&	WC	&	Para.	&	 14.0 	&	  $-35.645 $	&	bcc 	&	WC	&	Ferro.	\\
 6.8 		&	 $-49.249 $ 	&	sc 		&	WC	&	Para.	&	 14.5 	&	  $-34.880 $	&	bcc 	&	WC	&	Ferro.	\\
 7.0 		&	 $-48.919 $	&	sc 		&	WC	&	Para.	&	 15.0 	&	  $-34.142 $	&	bcc	&	WC	&	Ferro. 	\\
 7.2 		&	 $-48.566 $	&	sc 		&	WC	&	Para.	&	 15.5 	&	  $-33.432 $	&	bcc 	&	WC	&	Ferro. 	\\
 7.4 		&	 $-48.193 $	&	sc 		&	WC	&	Para.	&	 16.0 	&	  $-32.748 $	&	bcc	&	WC	&	Ferro.	\\
\end{tabular}
\end{ruledtabular}
\end{table*}

%*************************
\subsubsection*{2-jellium}
\begin{figure*}
\includegraphics[height=0.25\textheight]{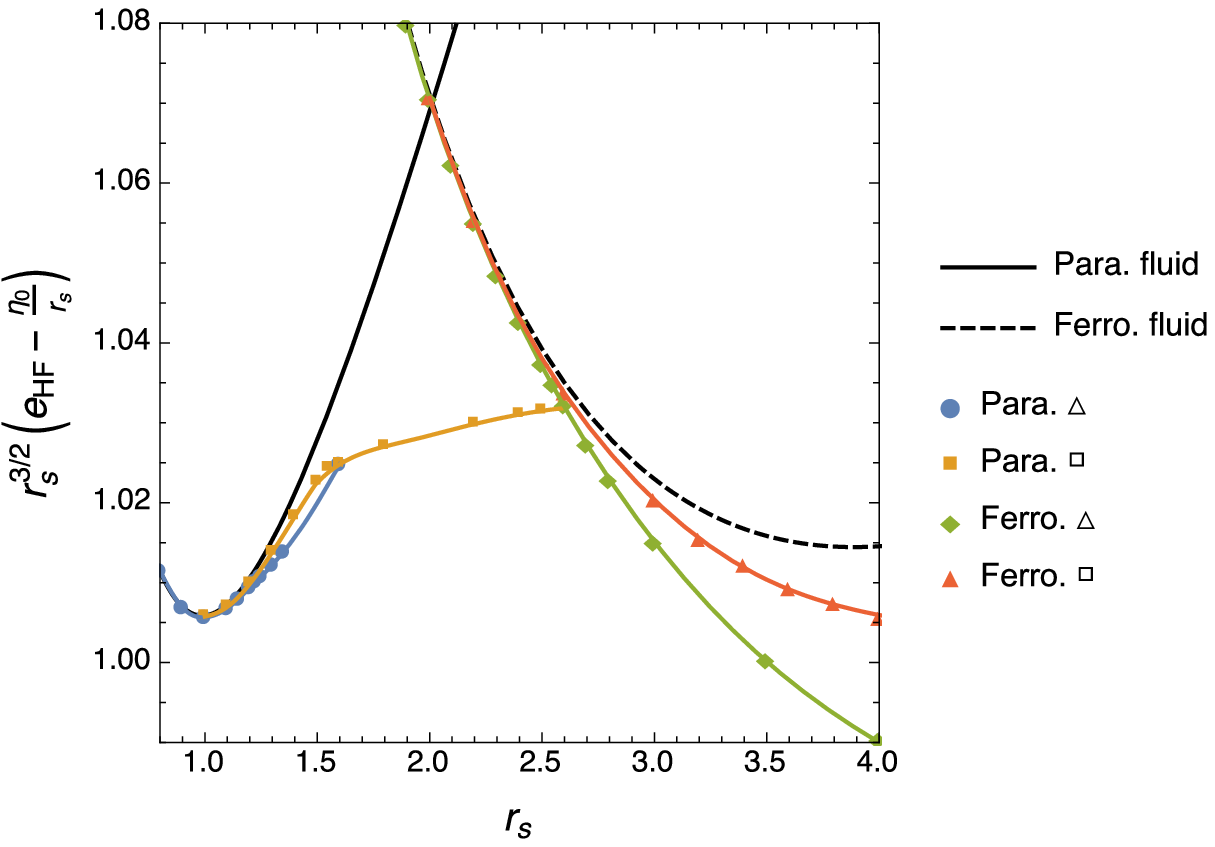}
\includegraphics[height=0.25\textheight]{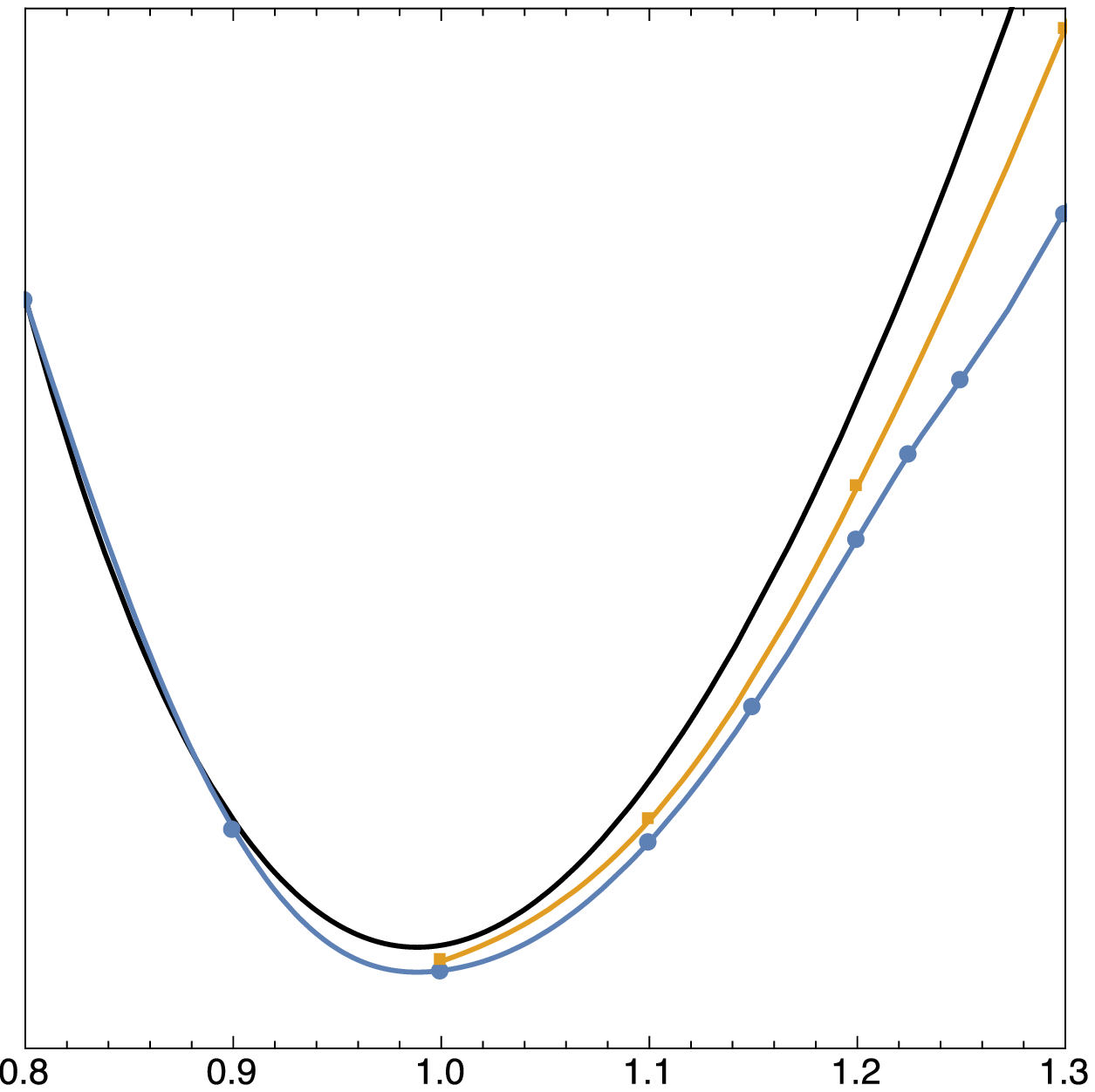}
\caption{
Left: SBHF phase diagram of 2-jellium constructed with the data of Ref.~\onlinecite{Bernu11}.
Right: SBHF in the high-density region ($0 < \rs < 1.3$).
\label{fig:2jellium-HF-PD}
}
\end{figure*}

%*************************
In 2D, Bernu et al. \cite{Bernu11} have obtained the SBHF phase diagram by considering the FF, the WC and the IC with square or triangular lattices. 
The phase diagram is shown in Fig.~\ref{fig:2jellium-HF-PD}. 
They have shown that the incommensurate phase is always favored compared to the FF, independently of the imposed polarization and crystal symmetry, in agreement with the early prediction of Overhauser about the instability of the FF phase. \cite{Overhauser59, Overhauser62}
The paramagnetic incommensurate hexagonal crystal is the true HF ground state at high densities ($\rs < 1.22$).
For $\rs > 1.22$, the paramagnetic incommensurate hexagonal crystal becomes a commensurate WC of hexagonal symmetry, and at $\rs \approx 1.6$ a structural transition from the paramagnetic hexagonal WC to the ferromagnetic square WC occurs, followed by a transition from the paramagnetic square WC to the ferromagnetic triangular WC at $\rs \approx 2.6$.
Interestingly, as at the DMC level (see Sec.~\ref{sec:QMC-2jellium}), they do not find a stable partially-polarized state.

%%% FIGURE 8 %%%

%*************************
\subsubsection*{1-jellium}
%*************************
To the best of our knowledge, the SBHF phase diagram of 1-jellium is unknown but it would probably be very instructive.

%----------------------------------------------------------------
\subsection{\alert{Finite-temperature calculations}}
%----------------------------------------------------------------
\alert{All the results reported in the present review concerned the UEG at zero temperature.
Recently, particular efforts have been devoted to obtain the properties of the finite-temperature UEG in the warm-dense regime using restricted path-integral Monte Carlo calculations. \cite{Brown13a, Filinov15, Schoof15}
The finite-temperature UEG is of key relevance for many applications in dense plasmas, warm dense matter, and finite-temperature DFT. \cite{Brown13b, Brown13c}}

%----------------------------------------------------------------
\section{
\label{sec:CCL}
Conclusion}
%----------------------------------------------------------------
\alert{Mark Twain once wrote, ``There is something fascinating about science. One gets such wholesale returns of conjecture out of such a trifling investment of fact.''  How true this is of the uniform electron gas!  We have no simpler paradigm for the study of large numbers of interacting electrons and yet, out of that simplicity, behavior of such complexity emerges that the UEG has become one of the most powerful pathways for rationalizing and predicting the properties of atoms, molecules and condensed-phase systems.  The beauty of this unexpected \emph{ex nihilo} complexity has lured many brilliant minds over the years and yet it is a siren song for, ninety years after the publication of Schr\"odinger's equation, a complete understanding of the UEG (even in the non-relativistic limit) continues to elude quantum scientists.}

\alert{In this review, we have focused on the energy of the UEG, rather than on its many other interesting properties.  We have done so partly for the sake of brevity and partly because most properties can be cast as derivatives of the energy with respect to one or more external parameters.  Such properties are attracting increasing attention in their own right and we look forward to comprehensive reviews on these in the years ahead.  However, we also foresee continued developments in the accurate calculations of the energies themselves.  These will play a critical role in the ongoing evolution of Quantum Monte Carlo methdology and will improve our understanding of, and our ability to model, phase transitions in large quantum mechanical systems.}

\alert{Many regard a full treatment of the uniform electron gas as one of the major unsolved problems in quantum science.  We hope that, by providing a snapshot of the state of the art in 2016, we will inspire the next generation to roll up their sleeves and confront this fascinating challenge.}
 
%----------------------------------------------------------------
\begin{acknowledgments}
%----------------------------------------------------------------
The authors would like to thank Neil Drummond, Mike Towler and John Trail for useful discussions, and Bernard Bernu and Lucas Baguet for providing the data for the phase diagram of 2-jellium.
P.F.L. thanks the Australian Research Council for a Discovery Early Career Researcher Award (Grant No.~DE130101441) and a Discovery Project grant (DP140104071).
P.M.W.G.~thanks the Australian Research Council for funding (Grants No.~DP120104740 and DP140104071). 
\end{acknowledgments}

\bibliography{Reviewium}

\end{document}